\documentclass[10pt,journal,compsoc]{IEEEtran}

\usepackage{amsthm,xpatch}
\usepackage{amsfonts}
\usepackage{xcolor}
\usepackage{tabularx}
\usepackage{graphicx}
\usepackage{booktabs}
\newtheorem{mydef}{Definition}
\newcommand{\indep}{\perp \!\!\! \perp}

\usepackage{enumitem}
\usepackage{multirow}

\ifCLASSOPTIONcompsoc
  \usepackage[nocompress]{cite}
\else
  \usepackage{cite}
\fi

\usepackage[numbers]{natbib}
\ifCLASSINFOpdf
\else
\fi
\usepackage{amsmath}
\begin{document}
%
\title{\huge Causal Inference for Recommendation: Foundations, Methods and Applications}
%
%
%
%
\author{Shuyuan Xu, Jianchao Ji, Yunqi Li, Yingqiang Ge, Juntao Tan, Yongfeng Zhang \thanks{The authors are with the Department of Computer Science, Rutgers University, USA. \\
Emails: \{shuyuan.xu, jianchao.ji, yunqi.li, yingqiang.ge, juntao.tan, yongfeng.zhang\}@rutgers.edu}}

\IEEEtitleabstractindextext{%
\begin{abstract}
Recommender systems are important and powerful tools for various personalized services. Traditionally, these systems use data mining and machine learning techniques to make recommendations based on correlations found in the data. However, relying solely on correlation without considering the underlying causal mechanism may lead to various practical issues such as fairness, explainability, robustness, bias, echo chamber and controllability problems. Therefore, researchers in related area have begun incorporating causality into recommendation systems to address these issues. In this survey, we review the existing literature on causal inference in recommender systems. We discuss the fundamental concepts of both recommender systems and causal inference as well as their relationship, and review the existing work on causal methods for different problems in recommender systems. Finally, we discuss open problems and future directions in the field of causal inference for recommendations.
\end{abstract}

}

\maketitle

\IEEEdisplaynontitleabstractindextext

%
\IEEEpeerreviewmaketitle

\IEEEraisesectionheading{\section{Introduction}\label{sec:introduction}}

\IEEEPARstart{R}{ecommender} systems have been recognized as one of the most effective tools to alleviate the information overloading, and have been widely deployed in many real-world systems, such as e-commerce platforms (e.g., Amazon, eBay), social networks (e.g., Facebook, Twitter), video-sharing platforms (e.g., Youtube, TikTok) and streaming services (e.g., Netflix, Hulu). 
In general, these systems use advanced techniques to learn users' preferences from historical data, along with collected user, item, and content information. And the development of these techniques has advanced rapidly in recent years.

Generally speaking, recommendation algorithms can be categorized into three major types: collaborative filtering, content-based recommendation and hybrid methods \cite{jannach2010recommender, zhang2019deep, adomavicius2005toward}. 
Collaborative filtering (CF) models are based on a key idea that similar users may share similar interest and similar items may be liked by similar users. Early memory-based CF models, such as user-based CF \cite{resnick1994grouplens,konstan1997grouplens} and item-based CF \cite{sarwar2001item,linden2003amazon}, take the row or column vectors of the user-item rating matrix as the user and item vector representations, and calculate the similarity between users or items for recommendation based on pre-defined similarity functions such as cosine similarity and Pearson correlation coefficient. To extract latent semantic meanings from the matrix, researchers later explored learned user and item vector representations. This started with Latent Factor Models (LFM) such as matrix factorization \cite{koren2009matrix}, 
probabilistic matrix factorization \cite{mnih2008probabilistic} and factorization machines \cite{rendle2010factorization}, which are widely adopted models in practice. In these models, each user and item is learned as a latent representation to calculate the matching score of each user-item pair, usually based on inner-product. The development of deep learning and neural networks has further extended CF models. For example, \cite{cheng2016wide,xue2017deep,he2017neural,hsieh2017collaborative} adopts simple user and item representations (e.g., one-hot vectors) and learns a complex matching function. \cite{zheng2017joint,zhang2017joint,zhang2016collaborative,ai2018learning,mcauley2015image} learn complex user and item representations and adopt a simple matching function (e.g., inner product). User representations can also be directly calculated from historical interactions, such as in sequential recommendation \cite{hidasi2016session,kang2018self}. Content-based recommendation will utilize rich information about users and items, or even context information, to enhance recommendation. In order to learn the similarities among items based on the side information, the representation approaches applied by the content-based recommendations have been developed from simple models such as TF-IDF \cite{kompan2010content} to deep learning based models such as DNN \cite{covington2016deep}, CNN \cite{liu2017deepstyle}, etc. Hybrid approaches combine collaborative filtering and content-based methods, which exploit the benefit of both methods and avoid their certain limitations \cite{zhang2019deep, burke2002hybrid, jannach2010recommender}.

The foundation of traditional recommendation algorithms is mining or learning the correlative pattern from data. For example, many collaborative filtering models aim to learn the user-item correlative pattern, some content-based recommendation models aim to learn the feature-feature correlative pattern. However, the real-world applications are driven by underlying causal mechanisms, pure correlative learning without considering the causation will lead to some practical issues. We take the classic ``beer and diapers'' problems as an example. Pure correlative learning will learn the strong correlation pattern between beer and diapers, thus recommend beer for customers bought diapers or vice versa. However, the underlying mechanism is that young fathers usually buy beer and diapers together, and recommending beer or diapers without considering the underlying mechanism will cause confusion and further hurt user's satisfaction. Therefore, it is important to advance from correlative learning to causal learning.

Formally speaking, causal inference studies the causal relation between cause and effect, where cause takes responsible for the effect. Two famous and popular frameworks are the potential outcome framework (also known as the Neyman–Rubin Potential Outcomes or the Rubin Causal Model) \cite{rubin1974estimating} and the structural causal model (SCM) \cite{pearl2009causality,pearl2009causal}. Both causal frameworks contribute to the development of causal recommendations. By leveraging the underlying causal mechanisms in recommender systems, causal recommendation is able to handle different practical issues, including explainability, fairness, robustness, uplift, and unbiasedness. 

\textbf{Contribution of this survey.} In this survey, we aim to provide a comprehensive review of causal inference for recommendation. We first introduce the fundamental knowledge of recommender systems and then discuss existing work of causal inference for recommendation. Specifically, we explore the causal inference in recommender systems in two dimensions. The first dimension follows the pipeline of causal inference, including concepts, notations, and techniques in causal inference, and the connection between causal inference and recommender systems. The second dimension follows the practical problems in recommendation, including problem introduction, causal methods, and open problems. More specifically, we include explainability, fairness, robustness, uplift-based, unbiasedness in recommendation. 
Finally, we highlight several open problems in causal inference for recommendation that remain to be addressed.

\textbf{Difference with Existing Surveys.} Several surveys in recommender systems or causal inference have been published in recent years. For example, \citet{zhang2020explainable} and \citet{chen2022measuring} review explainable recommendation, \citet{li2022fairness} and \citet{wang2022survey} review fairness in recommendation, \citet{ge2022survey}, \citet{wang2022trustworthy} and \citet{fan2022comprehensive} summarize trustworthy recommender systems, \citet{chen2020bias} review bias in recommendation, \citet{zhang2019deep} review the deep learning based recommendation algorithms, \citet{ko2022survey} provide a comprehensive review of recommender systems, \citet{yao2021survey} provide a comprehensive review of causal inference methods, \citet{guo2020survey} and \citet{vowels2021d} summarize existing methods on causal structural learning and causal discovery. \citet{gao2022causal} summarize existing work on causal inference in recommender systems. Unlike \citet{gao2022causal} mainly introduce existing work in perspective of recommender systems, our survey provide systematic review in perspective of both causal inference and recommender systems. 

\textbf{Organization.} This survey is organized as follows: Section \ref{sec:preliminary} introduces the preliminaries of recommender systems. From Section \ref{sec:causalnotations} to \ref{sec:causaldiscovery}, we introduce fundamental knowledge of causal inference and the connection with recommender systems. Section \ref{sec:explainability} to \ref{sec:unbiasedness} introduce existing causal methods on explainable recommendation, fairness in recommendation, uplift-based recommendation, robust recommendation, unbiased recommendation, respective. In Section \ref{sec:openproblems}, we discuss some open problems and future directions in causal inference for recommendation. Section \ref{sec:conclusion} concludes this survey.

\section{Preliminaries for Recommender Systems}\label{sec:preliminary}

In general, recommender systems aim to model user preferences based on collected information, including user profile, item profile, and user-item interactions, and further predict users' future interactions. User profile represents the registered information of the user, which may include user id, user age, user gender, user income, etc. Recommender systems may only use partial information for recommendation (e.g., using user id only). The term ``items'' represents different objects in differnt recommender systems (e.g., product in e-commerce, other users in social networks, videos in online video platform, etc.). According to different definition of ``item'', item profile may include different item features. For example, products in e-commerce may take brand, category, price, image , etc. in item profile; videos in online video platform item profile may take video length, content description, etc. in video recommendation; other users in social networks may take corresponding user profiles as item profile. Similarly, recommender systems may only partial information of item profile for recommendation. Interactions refer to possible user behaviors towards items according to defined task (e.g., click, purchase, rate, add-to-cart, review for e-commerce recommendation, like, dislike, share for video recommendation, etc.). In general recommender systems, interactions are typically represented in two ways, one is explicit feedback, the other is implicit feedback. Explicit feedback, such as ratings and reviews, is the explicit representation of users' preference (e.g., rating score as 5 means that user like this item), while implicit feedback, such as click, is collected during user-system interaction process and implicitly represent users' preference (e.g., user's click behavior means that it is likely that user likes the corresponding item).

Traditional recommendation algorithms can be roughly categorized into collaborative filtering, content-based recommendation and hybrid models. The basic idea of collaborative filtering (CF) is that similar users may share similar interests and similar items may be likede by similar users. CF methods can be further divided into memory-based CF and model-based CF. memory-based CF makes predictions by a simple similarity measurement over historical data. For example, user-based CF \cite{konstan1997grouplens,resnick1994grouplens} or item-based CF \cite{linden2003amazon,sarwar2001item} takes the row or column vector of the user-item rating matrix as the representation of each user or item and calculate the similarity by a simple measurement such as cosine similarity. Model-based CF leverage a model to learn the representation of users and items to make predictions. It starts from Latent Factor Models, such as matrix factorization \cite{koren2009matrix}, probabilistic matrix factorization \cite{mnih2008probabilistic}, tensor factorization \cite{karatzoglou2010multiverse}, etc. Deep learning and neural networks have further extend CF models. Deep CF methods can be further divided into similarity learning approach and representation learning approach. The similarity learning approaches \cite{zheng2017joint,zhang2017joint,zhang2016collaborative,ai2018learning,mcauley2015image} leverage simple representation of users and items (e.g., one-hot vectors) and learns a complex matching function to make prediction on each user-item pair. The representation learning approaches learn complex representation of users and items, and then apply a simple matching function (e.g., inner product) to calculate the prediction scores. Content-based recommendation \cite{van2000using, pazzani2007content,bobadilla2013recommender,alashkar2017examples,liu2017deepstyle,liang2015content,covington2016deep}, on the other hand, replies on rich user and item profile to recommend items similar to the ones the user prefered in the past. For example, in a movie recommender system, the model tries to understand the features (e.g., actors, directors, genres, tags, etc.) of movies that a user has rate highly in the past. Then, only the movies that match the preferred features of the user would be recommended. Hybrid models combine collaborative filtering and content-based methods, which exploit the benefit of both methods and avoid their certain limitations \cite{zhang2019deep, burke2002hybrid, jannach2010recommender, burke2007hybrid, ccano2017hybrid}. Moreover, several works, such as \cite{balabanovic1997fab, melville2002content}, have empirically demonstrated that the hybrid approaches are able to achieve more accurate recommendation than pure collaborative and content-based methods. 

Besides above traditional recommendation algorithms, there are some other recommendation algorithms. Sequential recommendation \cite{fang2020deep} (also related to session-based or session-aware recommendation), which leverage the timestamp information of interactions to suggest items, have become increasingly popular in academic research and industrial application. Traditional sequential recommendation models employ simple machine learning approaches to model sequential data, such as Markov chain \cite{rendle2010factorizing}, session-based KNN \cite{hu2020modeling}. With the development of deep learning techniques, many deep models obtain tremendous achievements in sequential recommendation, including RNN \cite{hidasi2015session}, LSTM \cite{wu2017recurrent}, CNN \cite{tang2018personalized, yuan2019simple}, attention models \cite{wang2018attention} and memory networks\cite{chen2018sequential}. Moreover, with increasing success achieved by foundation models (e.g., Large Language Models) on natural language tasks (e.g., T5 \cite{raffel2020exploring}, GPT-3 \cite{brown2020language}, OPT \cite{zhang2022opt}, PaLM \cite{chowdhery2022palm}), recommender system community, leverage the unique characteristic of recommender systems, has developed the research on personalized foundation models. For example, P5 \cite{geng2022recommendation}, as a pretrain, personalized prompt,and predict paradigm for recommendation, formulates recommendation as a language understanding and generation task to serve as a foundation model for many recommendation tasks.

The recommendation models learn users' preference based on collected information, and make recommendation based on learned preference. Specifically, a recommender system will provide a personalized recommendation list along with possible explanations to a specific user. Recommender systems will first predict user's preference towards a set of candidate items. Then the system will rank candidate items to provide personalized recommendation list. It is worth mentioning that the ranking process is not necessarily solely based on the predicted scores provided by the recommendation algorithm. It is possible to re-rank the list based on different demands, such as diversity, fairness, some business purpose, etc. After generating personalized recommendation list, some recommendation systems may provide explanations along with recommendations. The explanations can be either generated simultaneously with the recommendation or after the recommendation, depending on the recommendation model is explainable or black-box.

To evaluate the performance of recommender systems, it is important to define the characteristics of a good recommender system and quantify the characteristics. For a recommendation model with ability of predicting rating scores, a excellent model should be able to predict accurate ratings. Therefore, RMSE or MSE is used to evaluate the recommendation performance. By considering the accuracy of ranking list and whether the user's prefered items recommended by the list, some commonly used metrics include Precision, Recall, F-Measure, NDCG, ROC Curve, AUC, MRR, etc. Besides above metrics used to evaluate recommendation performance, some metrics are used to evaluate the recommendation model in perspective of other purpose. For example, Absolute Difference (AD) \cite{zhu2018fairness} is used to evaluate the fairness of recommender systems.

\vspace{-5pt}
\section{Causal Notations in Recommendation}\label{sec:causalnotations}
Causal inference is a critical research topic stemmed from statistics \cite{holland1986statistics,glymour2016causal,pearl2009causal}, and has been widely used in many domains for decades, such as computer science, public policy, economic, etc. In this section, we introduce causal notations and demonstrate how to apply them in recommendation.

\vspace{-5pt}
\subsection{What is Causation}\label{sec:causation}
Causation (also refer to as causality) is a terminology that is usually compared to and discussed with correlation. Although both correlation and causation explore the relationship between variables, it is well known that ``correlation does not imply causation'' \cite{glymour2016causal}. Causation takes a step further than correlation. Intuitively, causation explicitly applies to the case that event $A$ causes event $B$. On the other hand, correlation is a much simple relation that event $A$ is related to event $B$, but one event does not necessarily cause another event to happen. For example, a study has shown that the data of monthly ice cream sales is highly related to the number of monthly shark attacks across the United States. Although the two variables are highly correlated, it is impossible to conclude that consuming ice cream causes shark attacks (or vice versa). It is more likely that both ice cream sales and shark attacks increase in the summer due to other factors such as warm weather, which leads to both variables being correlated. Similar examples can be found in recommendations. The beer-and-diapers story is a good example to illustrate the difference between causation and correlation in recommendation. There is an observation that beer and diapers sell well together. Based on pure correlative learning, beer should be recommended for customers who bought diapers or vice versa because of the strong correlation pattern between beers and diapers. However, the underlying causal mechanism is that young fathers may pick up some diapers while buying beer. Therefore, directly recommending items without considering the underlying causation may lead to confusion and scarified recommendation performance. 
In general, understanding causation helps us to better understand how the world works and can improve the performance of recommendation systems.

To theoretically study the causation, it is required to understand the mathematical representation of causation. In general, there are two commonly used frameworks for causal inference, one is the potential outcomes framework (also known as the Neyman–Rubin Potential Outcomes or the Rubin Causal Model) \cite{rubin1974estimating} and the other is the structural causal model framework \cite{pearl2009causality,pearl2009causal} proposed by Pearl. Existing works usually introduce two framework separately, however, we think both frameworks are logically equivalent \cite{pearl2009causal} and follow the similar intuition. In the following sections, we will introduce those two frameworks following the intuitive idea of causation, including the connections and differences of two frameworks.

\subsection{Key mathematical notations of Causation.}

Causal inference refers to a process of drawing a conclusion that a specific \textit{treatment} was the ``cause'' of the \textit{outcome} that was observed \cite{frey2018sage}. 
In this case, the atomic goal is to estimate the outcome if any specific treatment has been applied. Both frameworks use mathematical notations to represent the desired value. For Rubin Causal Model, the basic element is called potential outcome.

\begin{mydef}
\textbf{(Potential Outcome)} A potential outcome is the outcome for an individual under a possible treatment
\end{mydef}

Let $X$ ($X\in \{x_1, x_2, \cdots, x_n\}$) denotes the treatment, where $n$ is the total number of possible treatments. Most of the literature considers the binary treatment, for example, taking medicine is denoted as $X=1$ and not taking medicine is denoted as $X=0$. Under the binary treatment, the group of individuals with treatment $X=1$ is named as the \textit{treated group}, and the group of individuals with treatment $X=0$ is called as the \textit{control group}. Generally, the potential outcome of treatment with value $x_i$ is denoted as $Y(X=x_i)$, which can be simplified as $Y(x_i)$. The average potential outcome of treatment with value $x_i$ can be denoted as $\mathbb{E}[Y(x_i)]$. For any individual, only one treatment can be applied while keeping other variables unchanged, thus only one potential outcome can be observed. Therefore, potential outcomes can be further divided into two categories, the observed one is named as observed outcome while the remaining unobserved potential outcomes are named as counterfactual outcomes.

In recommendation, the outcome is usually defined as user behavior (e.g., click, purchase) or user preference (e.g., rating). Unbiased recommendation models define the treatment as exposure, in which the observed feedback $Y$ (i.e., observed outcome) can be modeled as the product of two unobserved variables exposure $O$ and relevance $R$ (i.e., $Y=O\cdot R$) \cite{saito2020unbiased,saito2019unbiased,yang2018unbiased,zhu2020unbiased,ding2022addressing}. More specifically, in recommender systems, $Y=0$ can be either negative samples (i.e., $R=0$) or potential positive samples (i.e., $R=1$, $O=0$), which lead to data bias in recommendation. To achieve personalized recommendation, the models usually estimate the potential outcome $Y_{u,v}$ for a certain user-item pair $(u,v)$ (i.e., $Y_{u,v}=O_{u,v}\cdot R_{u,v}$). 
By correctly estimating potential outcome $Y_{u,v}(O_{u,v}=1)$ (i.e., $R_{u,v}=Y_{u,v}(O_{u,v}=1)$), the designed model is able to achieve unbiased recommendation. Uplift-based recommendation models define the treatment as recommendation (i.e., 1 for recommended, 0 for not recommended) \cite{sato2019uplift}. For each observed user-item pair, only one treatment can be observed (i.e., recommended or not recommended). Therefore, it is a challenge of estimating the counterfactual outcome to calculate the uplift value for recommendation. To achieve fairness, the treatment can also be defined as the sensitive attribute \cite{makhlouf2020survey}(e.g., 1 for privileged group and 0 for disadvantaged group).

Besides the treatment variable and the outcome variable, some other variables can be observed, and they can be further categorized as pre-treatment variables and the post-treatment variables \cite{yao2021survey}.

\begin{mydef}
(Pre-treatment Variables) Pre-treatment variables are the variables that are not affected by the treatment, which are also named as background variables.
\end{mydef}

\begin{mydef}
(Post-treatment Variables) Post-treatment variables are the variables that are affected by the treatment.
\end{mydef}

Different recommendation scenario may include different information and causal mechanisms, thus the specific definition of pre-treatment variables and post-treatment variables may vary. 

In addition to the potential outcome, Pearl uses another popular representation which distinguishes correlation and causation using $do$-operation \cite{pearl2009causality,glymour2016causal} from the perspective of probability. Supposed that $X$ denotes the treatment and $Y$ denotes the outcome, correlation and causation pursue different probabilities. Specifically, correlation estimates the conditional probability $P(Y|X)$ from observational data to determine the correlative relation between $X$ and $Y$. By contrast, causal inference estimates $P(Y|do(X=x_i))$ representing the outcome under a possible treatment $x_i$, where $do$-operation intuitively denotes applying the treatment instead of observing the treatment. The average outcome of applying treatment $x_i$ can be represented as $\mathbb{E}[Y|do(X=x_i)]$. A specific probability $P(Y=y|do(X=x))$ can be simplified as $P(y|do(x))$. As we mentioned before, existing causal frameworks follow the same intuition, therefore, the mathematical notations of $do$-operations and potential outcomes can be converted to each other in most cases. For example, in unbiased recommendation model, the treatment is usually defined as exposure. The results for a user-item pair $(u,v)$ under exposure can be expressed as $P(Y|u,v,do(X=1))$ where $Y$ is the outcome and $X$ is the exposure variable. If we define variable $V$ as exposed items, then it can also be represented by $P(Y|u,do(V=v))$. Similarly, the causal notations in uplift-based recommendation and fairness for recommendation can also be expressed by $do$-operations. 

By defining $do$-operations, \textbf{\emph{intervention}} as a basic concept in causal inference can be formally defined. As we mentioned above, the $do$-operation denotes applying the treatment, which can be also defined as the \emph{intervention} on the treatment variable. We will introduce more details in section \ref{sec:causalesti}. \textbf{\emph{Counterfactual}} is an important concepts in both the potential outcome framework and structural causal model, which represents the difference with factual. More specifically, counterfactual represents the scenario that the treatment variable had a different value compared with the observed value in the factual world. For example, considering the treatment as taking drugs and the outcome as recovery, a patient who took drugs and recovered may wonder if he would have been recovered if he hadn't taken the drugs. In this case, in the factual world, the patient took drugs and recovered, and in the counterfactual world, the patient did not take drugs and we wonder if he would recover. Similar example can be observed in recommender systems, for uplift-based recommendation, the treatment is defined as recommendation, the outcome is defined as user behaviors, and the system aims to maximize the increment of user behavior caused by recommendation. However, the item cannot be both recommended and not recommended in the factual world, therefore, it is necessary to apply counterfactual into recommendation. Counterfactual has been widely applied into recommender systems to address practical issues and made great success. We will demonstrate details in this survey.

\section{Causal Assumptions in Recommendation}\label{sec:causalassumption}
In this section, we will introduce commonly used assumptions in causal inference \cite{yao2021survey}. 

\begin{mydef}
\textbf{(Stable Unit Treatment Value Assumption (SUTVA))} The potential outcomes for any individual do not vary with the treatment assigned to other individual, and, for each individual, there are no difference forms or versions of each treatment level, which lead to different potential outcomes.
\end{mydef}

This assumption emphasizes the independence of each individual, which means that there are no interconnections between individuals. In recommendation, the individual usually represents the user. The traditional recommendation implicitly assumes the independence between users, which satisfies the SUTVA assumption. However, this assumption does not always hold in practical recommender systems. For example, in the recommendation for social networks, the users may connect with each other through the network structure. Some recommendation models do not have explicit users, for example, in session-based recommendation. In this case, the individual may be considered as the sessions, which temporally connect with each other.

\begin{mydef}
\textbf{(Ignorability)} Given the variables $W$, which are not affected by the treatment, treatment assignment X is independent to the potential outcomes, i.e., $Y(1), Y(0)\indep X|W$.
\end{mydef}

The ignorability assumption is also named as the unconfoundedness assumption. This assumption defines the treatment assignment under certain condition. Specifically, for individuals with the same variables $W$, the treatment assignment is random. This assumption is accepted by many recommendation algorithms, however, in real-world recommender systems, there may exists unobserved variables affect both the treatment and outcome, which has been studied by existing works \cite{xu2021deconfounded,wang2020causal}.

\begin{mydef}
\textbf{(Positivity)} For any value of variables $W$, which are not affected by the treatment, the treatment assignment is not deterministic:
\begin{equation}
    P(X=x|W=w)>0, \quad \forall x \text{ and  } w.
\end{equation}
\end{mydef}

This assumption guarantees the feasibility and significance of estimating the treatment effect. If for some values of $W$, the treatment assignment is deterministic, then the outcomes of at least one treatment can not be observed forever for these values. In this case, estimating the treatment effect is impractical and meaningless. This assumptions hold in recommendation algorithm design. For each user, every items have the chance to be exposed to the users. Items that cannot be exposed are not within the research scope of recommender systems.

With above three assumptions, the connection between the observed outcomes and the potential outcomes can be established.

\begin{equation}
\begin{aligned}
    \mathbb{E}[Y(x)|W=w] = &\mathbb{E}[Y(x)|W=w, X=x]\\
    =& \mathbb{E}[Y|W=w,X=x]\\
\end{aligned}
\end{equation}

Apart from above three commonly used assumptions, there is another way to represent the assumed mechanism. 
\begin{mydef}
\textbf{(Structural Causal Model (SCM))} A SCM consists of a set of endogenous ($V$) and a set of exogenous ($U$) variables connected by a set of functions ($F$) that determine the values of the variables in $V$ based on the values of the variables in $U$.
\end{mydef}

\begin{figure}
    \centering
    \includegraphics[width=0.5\textwidth]{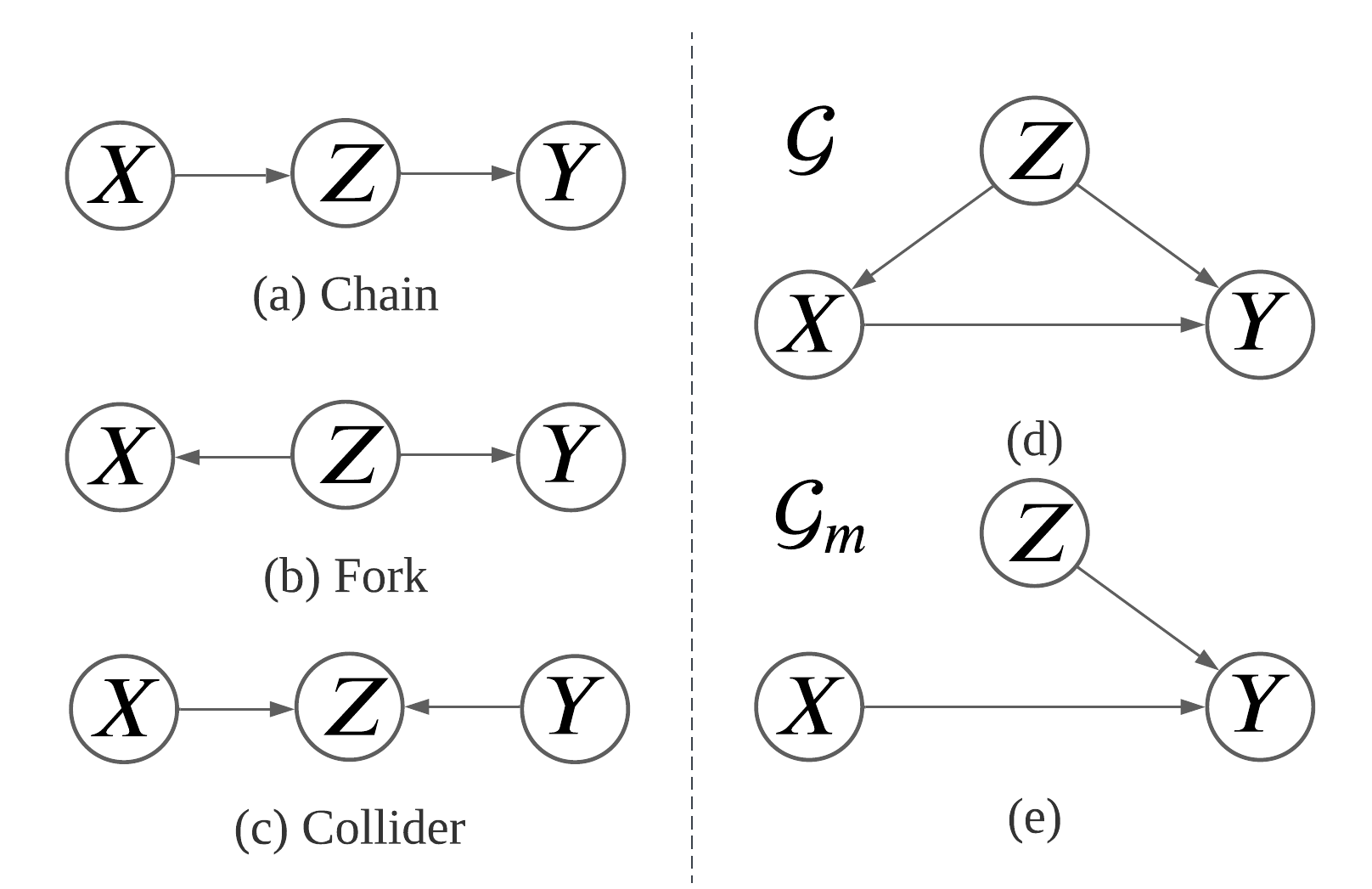}
    \caption{$X$, $Y$, $Z$ represent three variables. (a)-(c) show three fundamental causal graphs. (d) show and example of causal graph, and (e) represents the manipulated graph of (d) when intervene on variable $X$.}
    \label{fig:typicalDAG}
\end{figure}

SCM is the key concept in the Pearl's causal framework, which provides stronger assumptions (than potential outcomes framework) about the mechanisms behind the scenarios, which indicates the relationships between variables other than the treatment and the outcome. Each SCM is associated with a graphical model $\mathcal{G}$, represented as a Directed Acyclic Graph (DAG), where each node is a variable in $U$ or $V$ and each edge is a function $f$. Each edge corresponds to a causal assumption: If the variable $Y$ is the child of a variable $X$, then it is assumed that $X$ is the direct cause of $Y$; If the variable $Y$ is the descendant of a variable $X$, then it is assumed that $X$ is the potential cause of $Y$. The causal graph is the key difference between potential outcomes framework and structural causal model framework, where potential outcomes framework does not consider the causal graph to depict causal relationships. However, we think that both frameworks are built on some assumptions, and the causal graph is just a stronger assumption, which cannot completely separate two frameworks. We introduce three fundamental causal graph in Figure \ref{fig:typicalDAG}. 

\begin{figure}
    \centering
    \includegraphics[width=0.5\textwidth]{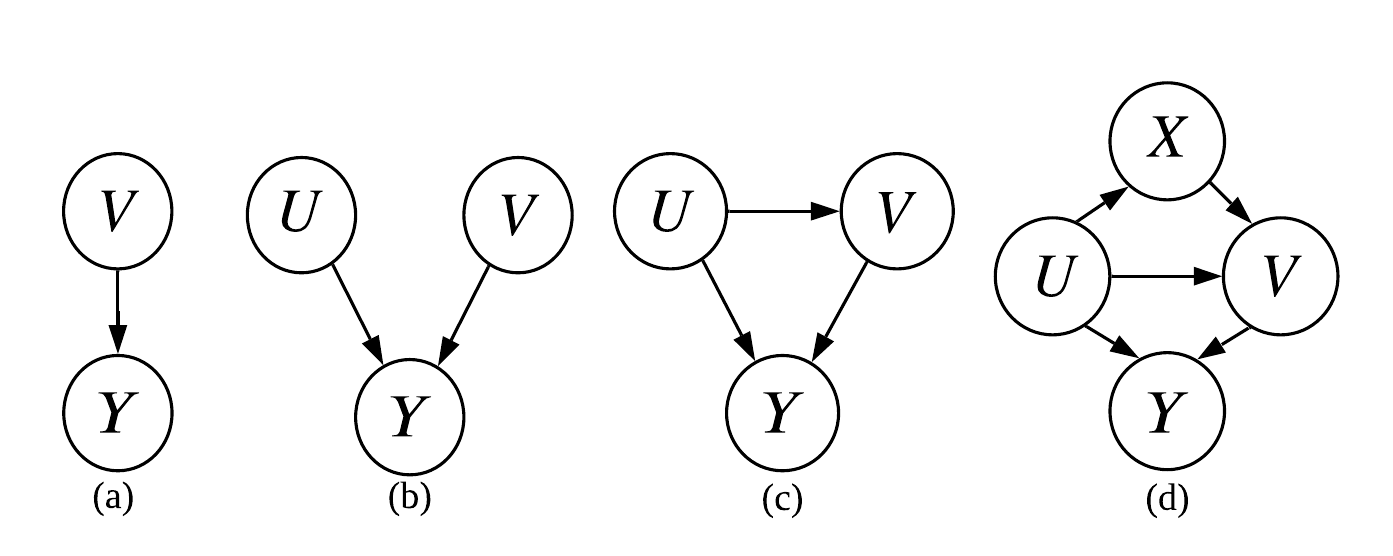}
    \caption{Many traditional recommendation and causal recommendation can be unified under different causal graphs. In the graphs, $U$ is user, $V$ is item, $X$ is user interaction history, $Y$ is preference score. (a) Causal graph for non-personalized models. (b) Causal graph for similarity matching-based CF models. (c) Causal graph that considers the causality from user to item \cite{bonner2018causal}. (d) Causal graph used in \cite{xu2021causal}.}
    \label{fig:ccf}
\end{figure}

Causal graph is a straightforward way to represents the underlying mechanism of recommender systems, and three typical causal graphs in Figure \ref{fig:typicalDAG} often appear in the mechanisms of recommender systems. For example, the chain structure in Figure \ref{fig:typicalDAG}(a) appears in \cite{xu2021deconfounded}, where item decides intrinsic item features and intrinsic item features further decides user preference; the fork structure in Figure \ref{fig:typicalDAG}(b) appears in \cite{zhang2021causal}, where item popularity is considered as a common cause of both item exposure and interaction probability; the collider structure in Figure \ref{fig:typicalDAG}(c) appears in \cite{zheng2021disentangling}, where user click is the common outcome of both user interest and conformity. For SCM, an existing work \cite{xu2021causal} has shown that the traditional recommendation and causal recommendation can be unified through a causal view, where the recommendation models aim to estimate $P(Y|U,do(X))$ (i.e., $Y$ represents the user preference, $U$ denotes users, $V$ denotes items) but with different causal graphs. More details can be found in Figure \ref{fig:ccf}.

As we mentioned before, the \emph{intervention} on the treatment variable can be interpreted as applying $do$-operation on the treatment variable. Intuitively, the $do$-operation means directly intervention, which cut off the influence from other variables to the treatment. Therefore, considering two variable $X$ and $Y$, the desired interventional probability $P(y|do(x))$ can be intuitively calculated as $P_m(y|x)$, which is the observed probability on the manipulated graph. Specifically, the manipulated graph removes all the income edges to the treatment variable. For example, considering a simple causal graph as Figure \ref{fig:typicalDAG} (d), where $X$ is the treatment, $Y$ is the outcome, and $Z$ is the confounder, $P(y|do(x))$ on the original causal graph $\mathcal{G}$ is the same as $P_m(y|x)$ on the manipulated graph $\mathcal{G}_m$ shown as Figure \ref{fig:typicalDAG} (e). An example in recommendation is taking intervention on item exposure, which generate the data of randomized experiments (i.e., data generation process follows the manipulated causal graph). We will introduce more details about randomized experiments in Section \ref{sec:causalesti}. Similar to the intervention on causal graphs, the intervention on structural equations take the intervened value as the input to calculate the output of the structural equations.

The introduced assumptions bridge the gap between the observed correlation and the estimated causation. We will introduce some commonly used methods based on introduced assumptions.

\section{Causal Effects in Recommendation}
After introducing the basic representation of the causal representation, many different kinds of causal effects can be defined using basic representations. 

A basic causal effect is called as the treatment effect (i.e., the outcome change if another treatment has been applied). More specifically, the treatment effect can be measured at the population, treated group, subgroup and individual levels. Here we define the treatment effect under binary treatment to make it clear, and it can be extended to multiple treatments by comparing their potential outcomes \cite{yao2021survey}. We takes the potential outcome as an example, and the $do$-operation can be applied in similar ways.

The treatment effect at the population level is named as the \textbf{Average Treatment Effect (ATE)} (some reference also name it as the \textbf{Average Causal Effect} \cite{glymour2016causal} or the \textbf{Total Effect} \cite{pearl2013mathematics}), which is defined as:

\begin{equation}\label{eq:ate}
    ATE=\mathbb{E}[Y(1)] - \mathbb{E}[Y(0)]
\end{equation}

The treatment effect at the treated group level is called \textbf{Average Treatment effect on the Treated Group (ATT)} (some reference also name it as \textbf{Effect of Treatment on the Treated (ETT)}\cite{glymour2016causal,pearl2009causality}), which is defined as:

\begin{equation}\label{eq:att}
    ATT=\mathbb{E}[Y(1)|X=1] - \mathbb{E}[Y(0)|X=1]
\end{equation}
where $Y(1)|X=1$ and $Y(0)|X=1$ represent the potential outcomes under both treatments of the treated group.

For the subgroup level, the treatment effect is named as \textbf{Conditional Average Treatment Effect (CATE)}, which is defined as:
\begin{equation}
    CATE = \mathbb{E}[Y(1)|W=w] - \mathbb{E}[Y(0)|W=w]
\end{equation}
where $W$ denotes the variables (i.e., grouping by multiple variables) defining the subgroup which are not affected by the treatment, and $Y(1)|W=w$ and $Y(0)|W=w$ are the potential outcomes under both treatments within the subgroup with $W=w$.

At the individual level, the treatment effect is called as \textbf{Individual Treatment Effect (ITE)}, which can be represented as:

\begin{equation}
    ITE = Y_i(1) - Y_i(0)
\end{equation}
where $Y_i(1)$ and $Y_i(0)$ are the potential outcomes for treatment $X=1$ and $X=0$ of individual $i$ respectively. The ITE is considered equivalent as the CATE \cite{shalit2017estimating,johansson2016learning} if each subgroup represents an individual. 

The treatment effect on different level has been used as quantitative evaluation in recommender systems to handle many issues. For example, ITE is used to estimate the uplift value of recommendation \cite{sato2019uplift,shang2021partially,xie2021causcf,sato2021causality}; ATE can be used to evaluate explanations \cite{xu2021learning} and estimate unbiased preference \cite{schnabel2016recommendations}; ATT is used to evaluate counterfactual fairness \cite{kusner2017counterfactual}; etc.

In addition to the treatment effect at different levels we introduced above, there are some causal effects for mediation analysis. A mediation model seeks to explain the process that underlines a causal relationship between the treatment and the outcome via the inclusion of a third variable, known as a mediator variable. Let $X$, $Y$, and $M$ denote treatment, outcome, and mediator respectively. We will introduce three types of effects under the binary treatment for mediation analysis.

First, Controlled Direct Effect (CDE) measure the expected increase in $Y$ as the treatment changes, while the mediator is set to a specific value $m$ for the entire population, which can be defined as:
\begin{equation}
\small
    CDE(m)=\mathbb{E}[Y|do(X=1,M=m)] - \mathbb{E}[Y|do(X=0,M=m)]
\end{equation}

Second, Natural Direct Effect (NDE) measures the expected increase in the outcome as the treatment changes, while the mediator is set to whatever value it would have attained prior to the change, i.e., $X=0$, which can be defined as:

\begin{equation}
\small
    NDE=\mathbb{E}[Y|do(X=1,M=M_0)] - \mathbb{E}[Y|do(X=0,M=M_0)]
\end{equation}
where $M_0$ represents the value of mediator under treatment as 0.

Finally, Natural Indirect Effect (NIE) measures the expected increase in outcome when the treatment is held constant at $X=0$, while $M$ changes to whatever value it would have attained under $X=1$, which can be defined as:
\begin{equation}
\small
    NIE = \mathbb{E}[Y|do(X=0,M=M_1)] - \mathbb{E}[Y|do(X=0,M=M_0)]
\end{equation}
where $M_1$ represents the value of mediator under treatment as 1. NIE captures the portion of the effect which can be explained by mediation alone.

The above direct and indirect effects play an important role in recommendation as well. The direct and indirect effects help the models quantitatively evaluate path-specific effects to detect and remove undesired effects. For example, they can be used to identify direct and indirect discrimination to achieve or explain fairness \cite{wu2018discrimination, zhang2018fairness}, they can be used to identify and remove some bias \cite{wang2021clicks,wei2021model}, etc.

\section{Causal Estimation Methods in Recommendation}
\label{sec:causalesti}
Having defined the causal effects, the next logical step is to ask how can we estimate those effects. One way is to perform a randomized experiment.

\subsection{Randomized Experiments.}
To measure the average treatment effect, an ideal way is to apply different treatment to the same group of individuals. However, the ideal solution is impractical in real-world situation. It can only be approximate by a randomized experiment. Specifically, a randomized experiment randomly assigns individuals into the treated group or the control group. The estimated ATE can be obtained by the difference of the average outcomes of two groups. To understand why a randomized experiment is the golden standard for estimating the average treatment effect, it is necessary to understand how correlation is different from causation. 

\begin{equation}\label{eq:observed}
    \begin{aligned}
        & \mathbb{E}[Y|X=1] - \mathbb{E}[Y|X=0] \\
        \overset{1}{=} & \mathbb{E}[Y(1)|X=1] - \mathbb{E}[Y(0)|X=0]\\
        \overset{2}{=} & \underbrace{\mathbb{E}[Y(1)|X=1] - \mathbb{E}[Y(0)|X=1]}_{\text{ATT}} \\
        + & \underbrace{\mathbb{E}[Y(0)|X=1] - \mathbb{E}[Y(0)|X=0]}_{\text{bias}}
    \end{aligned}
\end{equation}

Here, step 1 follows the fact that $Y(1)$ is the observed outcome when conditioning on $X=1$ and $Y(0)$ is the observed outcome when conditioning on $X=0$; step 2 adds and subtracts $\mathbb{E}[Y(0)|X=1]$ to construct the ATT term and the bias term. The bias term in Eq.\eqref{eq:observed} creates the gap between the correlation and causation. The randomized experiment eliminate the bias term by randomly assigning individuals into the treated group or the control group. More specifically, the random assignment makes the potential outcomes are independent from the treatment $Y(1),Y(0)\indep X$ (it does not imply that outcomes are independent from the treatment), thus $\mathbb{E}[Y(0)|X=1] = \mathbb{E}[Y(0)|X=0]$. Given $Y(1),Y(0)\indep X$, Eq.\eqref{eq:observed} can be rewrite as:

\begin{equation}
    \mathbb{E}[Y|X=1] - \mathbb{E}[Y|X=0] = \mathbb{E}[Y(1)] - \mathbb{E}[Y(0)]
\end{equation}

Therefore, a randomized experiment can simply estimate ATE as the difference of the average outcomes of the treated group and the control group. In recommendation, the randomized experiments are usually used to handle the bias \cite{rosenfeld2017predicting,bonner2018causal,yuan2019improving,liu2020general,jiang2019degenerate,yu2020influence,chen2021autodebias,lin2021transfer}. Specifically, by taking item exposure as the treatment, the randomized experiments follow the random recommendation policy instead of the deployed policy, in which return the unbiased data (i.e., also called as uniform data) for recommendation.

A randomized experiment is not a one-size-fits-all solution for causal inference. In reality, randomized experiments are always time-consuming and expensive, thus the study usually involve small number of individuals, which may not be representative of the population. Meanwhile, ethical concerns largely limit the applications of the randomized experiments such as environmental health studies. In addition, the randomized experiments cannot explain the causation on the individual level. Therefore, given the wide availability of observational data, the observational study is a shortcut for causal inference.

\subsection{Observational Data}
Although the observational study could be a shortcut for causal inference, there are some issues of the observational data should be carefully considered during designing the causal models. The existence of confounders is a critical problem in the observational data.

\begin{mydef}
\textbf{(Confounders)} Confounders are variables that affect both the treatment assignment and the outcome.
\end{mydef}

Due to the existence of confounders, some spurious effect may be observed (taking relationship between ice cream consumption and shark attacks as an example). Confounders widely exists in recommender systems. The existence of confounders often results in different bias based on the definition of confounders. For example, taking item popularity as a confounder, it will lead to popularity bias \cite{zhang2021causal}. In addition to some observed and measurable confounders, such as item popularity, some unobserved or immeasurable confounders (i.e., which violate the ignorability assumption in Section \ref{sec:causalassumption}) exist in real-world recommendation and have been widely studied by the community \cite{xu2021causal,wang2020causal,ding2022addressing}. 

\begin{table}[t]
    \centering
    \caption{Results of a study into a new drug, with gender being taken into account \cite{glymour2016causal}. a/b represents a out of b recovered.}
    \begin{tabular}{ccc}
    \toprule
         & Drug & No Drug \\\midrule
        Male & 81/87 (93\%) & 234/270 (87\%) \\
        Female & 192/263 (73\%) & 55/80 (69\%) \\
        Total & 273/350 (78\%) & 289/350 (83\%)\\\bottomrule
    \end{tabular}
    
    \label{tab:simpson}
\end{table}

Simpson's paradox is another phenomenon that could be observed in the observational data. From Table \ref{tab:simpson}, it can be observed that in both male and female groups, taking the drug has a better recovery rate; but in the total population, not taking drug has a better recovery rate. This phenomenon is usually caused by confounders. The Simpson's paradox can be also observed in recommender systems \cite{macdonald2021simpson}. \citet{macdonald2021simpson} observes the Simpson's paradox in offline evaluation for recommendation, and propose a method to mitigate the paradox in offline evaluation.

Compared to the experimental data, observational data only provides the information about what has occurred, but the why a specific treatment is token is unknown. Given that the treatment assignment mechanism is unknown, the bias term in Eq.\eqref{eq:observed} cannot be eliminated or quantitatively measured. Therefore, the bias caused by unknown treatment assignment is also a critical issue that should be carefully handled in model design.

\subsection{Methods Relying on Assumptions}
In some complex scenarios, it is risky to assume the causal mechanism based on prior knowledge. In this case, SUTVA, ignorability and positivity assumptions support some methods to estimate the potential outcomes.

One commonly used method is based on the idea of reweighting. As we mentioned before, due to the unknown treatment assignment mechanism, there may exists the bias problem. By assigning appropriate weight to each sample in the observational data, a pseudo-population can be created on which the distributions of the treated group and the control group are similar. There are two commonly used reweighting methods: inverse propensity scoring and confounder balancing.

\begin{mydef}
(Propensity Score) The propensity score is defined as the conditional probability of treatment given background variables:
\begin{equation}\label{eq:propensity}
    e(w)=P(X=1|W=w)
\end{equation}
\end{mydef}

Given the propensity scores defined above, inverse propensity scoring methods \cite{lunceford2004stratification,hirano2003efficient} assign a weight based on propensity score to each observed samples. Thus the estimated ATE based on the observed samples can be rewrite as:

\begin{equation}
    ATE_{IPS}=\frac{1}{n_1}\sum_{i,x_i=1}\frac{y_i}{e(w_i)} - \frac{1}{n_0}\sum_{j,x_j=0}\frac{y_j}{1-e(w_j)}
\end{equation}

 Inverse propensity scoring (IPS) is often used to design unbiased estimator for recommender systems \cite{schnabel2016recommendations,wang2021combating}, where the propensity score can be pre-defined or learned from the data.

Although the use of propensity score is effective to reduce the bias, there are some issues during applying IPS in practice. First, the correctness of the IPS estimator highly relies on the correctness of the propensity score estimator. To handle this dilemma, some augmented IPS methods are proposed, such as doubly robust estimator \cite{bang2005doubly}. Another drawback is that the IPS estimator has variance problem, that the estimator is unstable if the estimated propensity scores are small. To overcome this drawback, some methods propose to clip the propensity score \cite{saito2020unbiased} or trim samples with small propensity scores \cite{lee2011weight}.

Another reweighting method is confounder balancing \cite{hainmueller2012entropy,athey2018approximate,kuang2019treatment}. The motivation is that the confounders can be balanced by the moments, which uniquely determine the distribution of variables. Thus the sample weights can be learned to estimate the causal effect through reweighting. The confounder balancing based methods are used for stable learning \cite{shen2020stable} and robust recommendation \cite{li2022causal}.

In addition to reweighting methods, stratification is another representative method. The idea of stratification is to split the entire population into homogeneous subgroups, which makes the treated group and the control group are similar in each subgroup. Ideally, in this case, the samples in the same subgroup can be viewed as sampled from the data under randomized experiments. \citet{macdonald2021simpson} adopt this idea to mitigate Simpson's paradox in offline evaluation for recommendation.

In some applications, the causal mechanism is safely assumed based on prior knowledge or expert knowledge. In this case, the causal mechanism can be represented as a SCM as we introduced before. Although structural causal model framework requires stronger assumptions than potential outcomes framework, it also enable reasoning through the graph. Using a SCM, the key difference between causation and correlation is $do$-operations, which is the basic element to estimate the causal effect. As we mentioned, the $do$-operation can estimated by manipulated graph. However, the data from the manipulated graph is generated from randomized experiments. The approaches based on the data generated by the original causal graph are useful in practice. Applying backdoor adjustment is a popular approach.

\begin{mydef}
\textbf{(Back-door Criterion)} A set of variables $Z$ satisfies the backdoor criterion related to an ordered pair of variables $(X,Y)$ in a causal graph $\mathcal{G}$ if $Z$ satisfies both (1) No node in $Z$ is a descendant of $X$ and (2) $Z$ blocks every path between $X$ and $Y$ that contains an arrow into $X$.
\end{mydef}

Through identifying a set of variables satisfying the back-door criterion, the causal effect can be estimated using back-door adjustment formula.

\begin{mydef}
\textbf{(Back-door Adjustment)} If a set of variables $Z$ satisfy the back-door criterion related to an ordered pair of variables $(X,Y)$, and if $P(x,z)>0$, then the causal effect of $X$ on $Y$ is identifiable and is given by
\begin{equation}\label{eq:backdoor}
    P(y|do(x)) = \sum_z P(y|x,z)P(z)
\end{equation}
\end{mydef}

Given the population of the observed data, if we divide the subgroup based on value of $Z$, Eq.\eqref{eq:backdoor} can be considered as calculating the causal effect by the weighted sum of each subgroup, which is very similar to the stratification methods. Additionally, Eq.\eqref{eq:backdoor} can be rewrite as:
\begin{equation}
    P(y|do(x)) = \sum_z \frac{P(y,x,z)}{P(x|z)}
\end{equation}
where $P(x|z)$ is known as the ``propensity score'', therefore, the back-door adjustment is also an alternative representation of IPS methods. The back-door adjustment is widely used to address issues in recommendation, such as bias issues \cite{wang2021deconfounded, zhang2021causal}, echo chambers \cite{xu2022dynamic}, etc.

When we consider the $do$-operations, the interventions are not limited to actions that force a variable or a group of variables to take on specific value. In general, interventions may involve dynamic policies in which the treatment variable $X$ is made to respond in a specified way to some set $Z$ of other variables, which is denoted as $x=g(z)$. In this case, the estimated causal effect $P(Y=y|do(X=g(Z))$ can be calculated as:

\begin{equation}
    \begin{aligned}
        & P(Y=y|do(X=g(Z))) \\
        = & \sum_z P(Y=y|do(X=g(Z)),Z=z)P(Z=z|do(X=g(Z))) \\
        = & \sum_z P(Y=y|do(X=g(Z)),Z=z)P(Z=z) \\
        = & \sum_z P(Y=y|do(X=x),Z=z)|_{x=g(z)}P(Z=z)
    \end{aligned}
\end{equation}

In recommendation, the feedback data is collected from a deployed recommendation algorithm, thus the recommendation policy exists in the data generation process. Considering the dynamic policy as the recommendation policy, conditional intervention can also be applied to design causal recommendation models \cite{xu2021causal}. 
In recommendation scenario, observing an interaction in the feedback data does not imply that the interaction is destined to happen, thus the causal adjustment methods is sometimes applied with counterfactual reasoning \cite{xu2021causal,xu2021deconfounded,xu2022dynamic}. 

Apart from above adjustment formulas, there are some rules are valid for interventional probabilities, which are called as the rules of $do$-calculus. Before introducing the specific rules, we first introduce some notations. Let $X$, $Y$, $Z$, and $W$ be arbitrary disjoint sets of nodes in a causal DAG $\mathcal{G}$. $\mathcal{G}_{\overline{X}}$ denotes the graph obtained by deleting from $\mathcal{G}$ all arrows pointing to nodes in $X$. Likewise, $\mathcal{G}_{\underline{X}}$ denotes as the graph obtained by deleting from $\mathcal{G}$ all arrows emerging from nodes in $X$. The rules of $do$-calculus can be represented using above notations.

\begin{figure}
    \centering
    \includegraphics[width=0.5\textwidth]{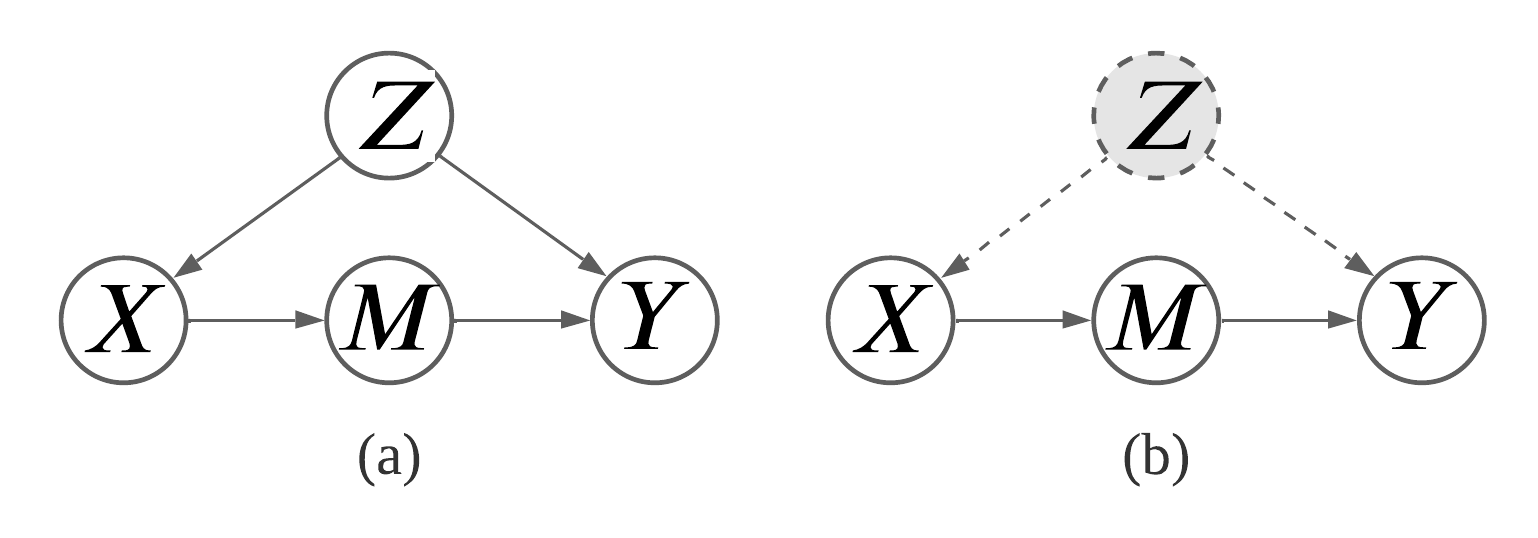}
    \caption{(a) An example of applying back-door adjustment on the causal graph. (b) An example of causal graph with an unobserved confounder, in which the causal values can be estimated by the front-door adjustment.}
    \label{fig:causalAdj}
\end{figure}

\begin{mydef}
\textbf{(The rules of $do$-calculus)} The following three rules are valid for every interventional distribution compatible with a causal graph $\mathcal{G}$
\begin{description}
\item[\textbf{Rule 1}] (Insertion/deletion of observations):
\begin{equation}
\begin{aligned}
    & P(y|do(x),z,w)=P(y|do(x),w) \\
    & \text{if} \quad (Y\indep Z|X,W)_{\mathcal{G}_{\overline{X}}}
\end{aligned}
\end{equation}
\item[\textbf{Rule 2}] (Action/observation exchange):
\begin{equation}
\begin{aligned}
    & P(y|do(x),do(z),w)=P(y|do(x),z,w) \\
    & \text{if} \quad (Y\indep Z|X,W)_{\mathcal{G}_{\overline{X}\underline{Z}}}
\end{aligned}
\end{equation}
\item[\textbf{Rule 3}] (Insertion/deletion of actions):
\begin{equation}
\begin{aligned}
    & P(y|do(x),do(z),w)=P(y|do(x),w) \\
    & \text{if} \quad (Y\indep Z|X,W)_{\mathcal{G}_{\overline{X}\overline{Z(W)}}}
\end{aligned}
\end{equation}
where $Z(W)$ is the set of $Z$-nodes that are not ancestors of any $W$-nodes in $\mathcal{G}_{\overline{X}}$.
\end{description}
\end{mydef}

With the help of the rules of $do$-calculus and introduced adjustment formulas, the interventional probabilities can be estimated by the observational data.

\subsection{Methods with Relaxed Assumptions}
Although above methods relying on introduced assumptions basically satisfy the requirement of estimating causal effect from the observational data, in practice, for some specific applications, the introduced assumptions may not always hold. There are some methods trying to estimate the causal effect with relaxed assumptions. 

SUTVA assumes that individuals are independent and identical distributed. However, in some real-world applications, such as social networks, SUTVA cannot hold anymore since individuals are inherently interconnected with each other through the network structure. To handle this issue in real applications, a commonly used approach is applying a model, which capture the interconnection, into a causal inference model. For examples, applying graph convolutional networks into a causal inference model to handle the network data \cite{guo2020learning}. 

The ignorability assumption assumes that the treament assignment is independent to the potential outcomes given the background variables. However, it is impossible to identify and collect all the background variables in real world, thus the ignorability assumption is hard to satisfy. In other words, there may exist unobserved confounders as we mentioned before. Only using observational data to estimate the causal effect is difficult, an alternative way is to combine the limited experimental data and observational data together \cite{kallus2018removing}. In recommendation, the unbiased data is collected from randomized experiments, using a small part of unbaised data and a large part of observed feedback is a popular way to design unbiased recommendation models. 

Another solution is based on the assumed SCM, which models the unobserved confounders into the causal graph (an example is shown in Figure \ref{fig:causalAdj}(c)). Similar to applying the back-door adjustment, we first identify a set of variables satisfying the front-door criterion.

\begin{mydef}
\textbf{(Front-door Criterion)} Given an ordered pair of variables $(X,Y)$ in a causal graph $\mathcal{G}$, a set of variables $Z$ satisfies the front-door criterion with respect to $(X,Y)$ if $Z$ satisfies the following conditions:
\begin{itemize}
\renewcommand\labelitemi{--}
    \item $Z$ intercepts all directed paths from $X$ to $Y$.
    \item There is no unblocked back-door path from $X$ to $Z$.
    \item $X$ blocks all back-door paths from $Z$ to $Y$.
\end{itemize}
\end{mydef}

Given a set of variables that satisfies the front-door criterion, we can identify the causal effect with unobserved confounders \cite{glymour2016causal}. 

\begin{mydef}
\textbf{(Front-door Adjustment)} If a set of variables $Z$ satisfy the front-door criterion related to an ordered pair of variables $(X,Y)$, and if $P(x,z)>0$, then the causal effect of $X$ on $Y$ is identifiable and is given by
\begin{equation}\label{eq:frontdooradj}
    P(y|do(x)) = \sum_z P(z|x)\sum_{x^\prime}P(y|x^\prime, z) P(x^\prime)
\end{equation}
\end{mydef}

The existence of unobserved confounders is widely recognized by the community \cite{xu2021deconfounded,zhu2022mitigating,ding2022addressing,wang2020causal}, there are some works \cite{xu2021deconfounded,zhu2022mitigating} that attempt to apply front-door adjustment in recommendation.

Using instrumental variables is a possible way to get around the ignorability assumption and conduct causal inference. Instrumental variables are defined as variables that only affect the outcome via the treatment variables. Typical instrumental variables methods \cite{angrist2009mostly,hartford2017deep} adopt two-stage models: the first stage reconstructs the treatment variable based on the instrumental variable and the second stage reconstructs the outcome based on the treatment from the first stage. In recommender systems, \citet{si2022model} adopt the instrumental variable to design a model-agnostic recommendation framework using search data.

\section{Causal Discovery in Recommendation}\label{sec:causaldiscovery}
The above methods aim to learn the causal effect, there is another branch of causal models targeting at learning causal relations, which is also known as causal discovery.
Except for few works only aim to identify treatment and outcome \cite{bengio2019meta}, most of the works aim to discover causal graphs. Following \cite{malinsky2018causal,guo2020survey}, traditional methods can be divided into three categories: constraint-based, socre-based and those based on functional causal models.

Constraint-based Algorithms learn a set of causal graphs that satisfy the conditional independence embedded in the data and statistical tests are utilized to verify if a candidate graph satisfies the independence. Score-based algorithms learn causal graphs by maximizing the scoring function $S(\mathbf{X}, \mathcal{G})$, which returns the score of the causal graph $\mathcal{G}$ given data $\mathbf{X}$. Algorithms based on Functional causal models (FCMs) usually define a variable as a function of its directed causes and some noise term (e.g., linearly weighted by the adjacency matrix of the causal graph \cite{shimizu2006linear}) and optimize the designed objective to learn the parameters of the functions. We only briefly introduce the causal discovery methods, interested readers may refer to \cite{malinsky2018causal,guo2020survey} for more details.

Most existing works in causal recommendation are based on pre-defined causal graph representing the underlying causal mechanisms. The pre-defined causal graphs are usually defined based on expert knowledge, which may be inaccurate and quite simple (i.e., only involve few variables). Leveraging causal discovery in recommendation will handle these issues. There exist few works \cite{wang2022sequential,xu2022causal} design recommender systems with causal discovery techniques based on continuous optimization \cite{zheng2018dags}. The learned causal mechanism will increase the explainability of recommender systems and guide the model design for other aspects, such as fairness, unbiasedness.

\section{Causal Explainability in Recommendation}\label{sec:explainability}
With the development of machine learning, accuracy is no longer the only only pursuit. Moreover, transparency and trustworthiness start to obtain increasing attention. For example, heathcare AI is required to provide not only accurate diagnoses, but also supporting explanations to convince patients. Recommender systems, with humans in the loop, also require transparency. Explainable recommendation, which emerged and developed with the pursuit of transparency and trustworthiness of recommender systems, has been increasing popular in both academia and industry. It aims to provide explanations for the recommended items, which will benefit the community in many ways. For consumers, the explainable recommendation is able to help them make better decisions. For the platform, it may improve the transparency, persuasiveness, trustworthiness and user's satisfaction of the system. For model developers, it is an important tool to understand the designed model and accelerate the design cycle. In this section, we will first introduce the overview of the explainable recommendations, and then summarize the existing causal methods, as well as some open problems related to causal inference.
\subsection{Problem Introduction}
The research of explainable recommendation, as a sub-area of explainable AI, was proposed and defined by \cite{zhang2014explicit}. With the rapid development of deep neural networks, the state-of-the art recommender systems widely adapt deep models to improve the recommendation performance. However, these deep models are too complicated for users to understand the decision made by the intelligent systems, thus a deep model is usually considered as a black-box. Recommender systems, serve as essential decision-making systems in daily life, are required to provide accurate decision results as well as underlying reasons. For example, a stock investor needs to know which characteristics lead to the recommendation before making the final decision. A consumer hopes to understand why the recommended items are worth buying before paying.

The explainable models can be either model-intrinsic or model-agnostic. The former one refers to generating explanations simultaneously with the recommendation results and the later one refers to generating explanations after providing the recommendation results. Model-intrinsic (also known as ad-hoc) explainable models usually design the explanation generation mechanism as a part of decision-making process, and model-agnostic explanations (also known as post-hoc) explainable models usually design separate mechanisms for generating explanations.

The explanations can be presented in many different ways, which usually depend on what kind of information source is used for explanations. Typically, the explanations can be presented as related users or items \cite{peake2018explanation,xu2021learning}, the features of users or items \cite{zhang2014explicit,tan2021counterfactual}, generated textual sentence \cite{wang2018reinforcement,li2021personalized}, visual explanations \cite{geng2022improving}, graph \cite{xian2019reinforcement}, etc. Existing works have made many successes in explainable recommendations with different information sources. For example, \citeauthor{zhang2014explicit} \cite{zhang2014explicit} propose Explicit Factor Model (EFM) which extract explicit item features and user opinions from user reviews to provide feature-level explanations. \citeauthor{peake2018explanation} \cite{peake2018explanation} extract association rules to provide purchased items as an explanation in a model-agnostic manner. \citeauthor{xian2019reinforcement} \cite{xian2019reinforcement} perform explicit reasoning path with knowledge graph to provide recommendations and explanations. In addition, Existing works have introduced explainability into conversational recommendation. \citeauthor{chen2021towards} \cite{chen2021towards} develop an Explainable Conversational Recommendation (ECR) model to provide accurate recommendations as well as high quality explanations by multi-round conversations. Incorporating causal inference ideas and techniques brings new opportunities for explainable recommendations. In the following part, we will focus on causal-related methods. Interested readers may refer other surveys \cite{zhang2020explainable,chen2022measuring} for more explainable recommendation approaches. 

\subsection{Causal Methods}

\subsubsection{Counterfactual}
In recent years, counterfactual reasoning draws more and more attention in explainable AI. For any AI system that makes predictions based on machine learning models, no matter white-box or black-box, counterfactual reasoning looks for what input (e.g., aspects, features) should be changed, and by how much, to acquire a different prediction. Then, the altered input will comprise the explanation. For instance, when generating explanations for a rejected loan request, it could be something like: if your annual income is $50,000$, instead of $30,000$, your request will not be rejected. Some existing works have introduced the idea of counterfactual reasoning into recommendation scenarios for generating explanations, which looks for minimal changes in the recommender system (e.g. item features, items in the history, user's behaviors, etc.) leading a different prediction to identify the most essential part (e.g. item features, items in the history, user's behaviors, etc.) as the explanations. \citeauthor{ghazimatin2020prince} \cite{ghazimatin2020prince} generate explanations for a recommender system based on users' actions in in the history. More specifically, it introduces a searching algorithm on a knowledge graph to look for the minimal set of user's history to be cut off, such that the user will receive different recommendation results. 
\citeauthor{tan2021counterfactual} \cite{tan2021counterfactual} proposes a counterfactual explanation framework for generating feature-level explanations. It introduces two new concepts, explanation complexity and explanation strength. These two concepts are used to formulate a counterfactual optimization problem, as well as an evaluation metric to evaluate the generated explanations. Later in \cite{ge2022explainable}, a similar counterfactual explanation framework is also used to explain which features are causing fairness issues in recommender system. \citeauthor{tran2021counterfactual} \cite{tran2021counterfactual} utilizes an influence function to analyze the training data. Then, a counterfactual set of training data are used for generating explanations. 

\subsubsection{Causal Discovery}
Explainable recommendation models based on causal discovery are still in theirs infancy. Causal discovery methods aim to extract causal relations among variables from the data. Existing causal discovery based approaches in recommendation provide model-intrinsic explanations. More specifically, through the extracted causal relations, the causal discovery based recommendation models are able to provide recommendations simultaneously with corresponding causal relations as explanations. As we mentioned, causal discovery methods usually try to learn a causal graph. In recommendation scenario, considering the extremely large amount of items, the learned causal graphs are typically based on item group level. For example, \citet{wang2022sequential} propose to learn a cluster-level causal graph to guide sequential recommendation. Based on the learned cluster-level causal graph and cluster assignment for each item, the model is able to calculate the causal relations between items. The item in interaction history with the strongest causal relation with the recommended item is identified as the explanation. \citet{xu2022causal} aim to learn a causal graph on product type (PT) level for PT-level recommendation. Particularly, the model takes collected feedback data as the result of the mixture of two completing mechanisms: a causal mechanism based on user intention and a intervention mechanism based on deployed recommendation algorithm. The recommendation and corresponding explanations are generated via the learned PT-level causal graph.

\subsection{Open Problems}
Despite the above successful usages in causal explainable recommendation, there are open problems that expected to be solved in the future. First, causal discovery based explainable recommendation models, are capable of generating model-intrinsic explanations, need further exploration. Second, the current counterfactual explanation algorithms are claimed to be model-agnostic because they are able to be applied on any recommendation models (or at least a wide range of recommendation models). However, the model itself has to be reachable. It is not certain about how to apply counterfactual explanation algorithms on an recommendation model that are not accessible by the algorithm user. Finally, there are currently no methods to leverage other causal reasoning methods, such as the do-calculus, to generate explanations.

\section{Causal Fairness in Recommendation} \label{sec:fairness}
Recommender system, as a powerful tool for business, has been widely used to improve user engagement and further create higher profit. Classical recommender systems mainly care about how to precisely estimate user preferences. However, in recent years, concerns about fairness in recommendation have attracted much attention from both industry and academia \cite{li2021towards,li2021user,li2021tutorial,li2022fairness,wang2022survey}. 
With the development of recommendation techniques, recommender systems have been widely used to assist or even replace human decision-making in several domains. Several studies have shown that the unfairness may lead to negative consequences \cite{mansoury2020investigating,d2020fairness,abdollahpouri2019multi}, which in turn may have significant social impacts. For example, in e-commerce, the unfairness of exposure of items may hurt the benefits of the platform and providers in long-term \cite{ge2021towards}; in educational recommendation \cite{dascalu2016educational}, an unfair system due to gender imbalance \cite{beede2011women} may discourage females from selecting STEM (i.e., science, technology, engineering, and mathematics) topics, which may affect society for generations; An unfair ad recommendation may even result in racial discrimination \cite{sweeney2013discrimination}. Therefore, to increase the applications of recommender systems and maintain a healthy social impact, it is critical to consider fairness in recommendations and build a reliable decision-making system.

\subsection{Problem Introduction}
Before achieving fairness in recommender systems, one should first understand the reasons of unfairness. Bias and discrimination are two commonly accepted causes of unfairness \cite{mehrabi2021survey,li2022fairness,ge2022survey,wang2022survey}. Biases in recommender systems mainly consist of bias in data and bias in algorithm. The bias in data may come from data generation, collection, sampling, and storage. For example, in recommender systems, the training data is collected from a deployed system, if the algorithm underlying the deployed system makes biased predictions, then the generated data may involve biases. The bias in data may affect the algorithms, since most machine learning algorithms rely on data to be trained and make predictions after training. If the training data contains biases, the algorithms trained on them will learn biased knowledge from these biases and further lead to unfairness. For example, if the training data shows significant imbalance between majority user/item group and minority user/item group, it is high likely that the recommendation algorithm learns much better on the majority group and results in discrimination on the minority group. Except for the bias in data, the recommendation algorithm itself may enhance existing biases and cause unfairness, which is referred to the bias in algorithm. For example, some recommendation algorithms may enhance the popularity bias, where popular items will get more recommendation than less popular items with equal or similar quality. 
Discrimination, as a multidisciplinary problem \cite{marshall1974economics,willborn1984disparate,romei2014multidisciplinary}, is also a cause of unfairness defined as an unjustified difference in treatment on the basis of any physical or cultural characteristic (e.g., race, gender, etc.) due to human prejudice and stereotyping. It is worth mentioning that unfairness is not only caused by bias and discrimination. For example, there may exist conflicts or trade-offs between different kinds of fairness \cite{kleinberg2016inherent,li2022fairness,ge2022survey}, where achieving one fairness will hurt another fairness. 

To fight against unfairness, it is important to define fairness. In general machine learning, fairness can be defined on target level (i.e., to achieve fairness on group-level or individual-level). Specifically, fairness can be categorized into group fairness and individual fairness.
\begin{itemize}[leftmargin=*]
    \item \textbf{Group Fairness}: Group fairness defines the fairness on group-level, which is based on the idea that different groups should be treated equally. Here the groups can be divided in many ways, where the most commonly used way is to split the groups based on some explicit sensitive attributes. 
    \item \textbf{Individual Fairness}: Individual fairness defines fairness on individual-level, which is based on the idea that similar individuals should receive similar predictions. Moreover, individual fairness can be theoretically considered as a very special group fairness, which divides each individual into different groups. 
\end{itemize}

Since fairness in recommender systems relates to the benefits from multiple stakeholders \cite{abdollahpouri2019multi,mansoury2021fairness,burke2017multisided,gharahighehi2021fair,wu2022joint}, the request of fairness may come from different sides. Therefore, the definition of fairness in recommendation can also be divided into user-side fairness and item-side fairness.

\begin{itemize}[leftmargin=*]
    \item \textbf{User-side Fairness}: User-side fairness aims to satisfy the fairness requirements from users (consumers). The request from the user side are mainly focusing on the recommendation quality (i.e., recommendation performance). The user-side fairness can be achieved on both group-level and individual-level. User-side fairness on group-level aims to reduce the discrepancy of recommendation quality between different user groups, where the user groups are divided by sensitive features, such as race or gender \cite{yao2017beyond,mansoury2020investigating}, or by assigned features (e.g., cold users vs. heavy users \cite{wu2022big}, active users vs. inactive users \cite{li2021user,fu2020fairness}). For user-side fairness on individual-level, the recommendation quality should be unchanged even an individual's sensitive features have changed. For example, \citet{li2021towards} incorporate the idea of counterfactual fairness \cite{kusner2017counterfactual} to design a recommendation model which makes the recommendation performance unchanged even the user's sensitive features are flipped in the counterfactual world.
    
    \item \textbf{Item-side Fairness}: Item-side fairness aims to satisfy the fairness from items side, which mainly focuses on requesting equal exposure opportunity of items to maintain market fairness. Here the items refer to ``items'' to be ranked or recommended. For example, in e-commerce, the items refer to products to be sold; in recruitment system, the items refer to job seekers (item providers). One branch of existing work focuses on achieving fairness according to item attributes. For example, some works \cite{kamishima2014correcting,abdollahpouri2017controlling,ferraro2019music,abdollahpouri2019unfairness,abdollahpouri2019managing,ge2021towards} achieve the fair exposure between popular and unpopular items to prevent unpopular items from being under-exposed. Moreover, another branch of research work mainly focuses on achieving fairness based on the sensitive attributes of item providers, such as gender \cite{boratto2021interplay,fabbri2020effect,ekstrand2018exploring}, geographic provenience \cite{gomez2021winner,gomez2021disparate,gomez2022provider}, etc.
\end{itemize}

It is worth noting that the user-side fairness and item-side fairness may not exclusive to each other, where two-sided fairness \cite{wu2021tfrom,chakraborty2017fair,patro2020fairrec,suhr2019two,wang2021user} approaches are proposed to satisfy the fairness demands from both sides. Besides the taxonomies mentioned above, there are also some taxonomies \cite{ge2022survey,li2022fairness} that are used to classify fairness in recommendation from other perspectives. For example, static fairness vs. dynamic fairness \cite{d2020fairness,creager2020causal,d2020fairness,williams2019dynamic}; short-term fairness vs. long-term fairness \cite{zhang2020fair,ge2021towards}; populational fairness vs. personalized fairness \cite{li2021towards,wu2022selective,bose2019compositional}; blackbox fairness vs. explainable fairness \cite{ge2022explainable}, centralized fairness vs. decentralized fairness \cite{liu2022fairness,maeng2022towards}. 

Typically, the proposed approaches to achieve fairness in recommendations can be roughly divided into three categories: pre-processing methods, in-processing methods and post-processing methods \cite{caton2020fairness,li2022fairness,ge2022survey,mehrabi2021survey}. Pre-processing methods usually aim to achieve fairness by minimizing the bias in the data before the model training. Compared with other types of methods, there are fewer works on pre-processing methods. Some representative methods include fairness-aware data sampling approach to cover items of all groups, data balancing approach \cite{ekstrand2018all} to increase the coverage of minority groups and data repairing approaches to ensure label correctness and remove disparate impact \cite{gao2021addressing}. In-processing methods propose to incorporate fairness requirements as a part of the objective function to achieve fairness during the training. Typically, the fairness requirement works as a regularizer or a constraint \cite{abdollahpouri2017controlling,beutel2019fairness,ge2021towards,li2021user,burke2017balanced,farnadi2018fairness,xiao2017fairness,yao2017beyond,yao2017new,zhu2018fairness}. To minimize the unfairness while minimizing the original loss function (i.e., recommendation accuracy loss), it is also important to find a trade-off between recommendation accuracy and fairness \cite{ge2021towards,ge2022toward}, which is also sometimes formulated as a multi-objective learning problem \cite{ge2022toward}. Post-processing methods aim to achieve fairness in inference stage after the training, by techniques such as re-ranking \cite{li2021user,singh2018fairness,yang2021maximizing,zehlike2017fa} or multi-armed bandit \cite{celis2019controlling,joseph2016fair,chen2020fair}. To measure the unfairness, many different fairness metrics are proposed. For example, Absolute Difference (AD) \cite{zhu2018fairness} measures the absolute difference between the performance of protected group and unprotected group; Normalized Discounted KL-divergence \cite{geyik2019fairness} calculates a normalized discounted cumulative value of KL-divergence for each position, etc. More possible fairness metrics can be found in \cite{wang2022survey}.

Recently, researchers have noticed that fairness cannot be well detected by solely correlation or association. Specifically, fairness criteria are based on solely joint distribution of random variables \cite{khademi2019fairness}, such as outcomes, features, sensitive attributes, etc. However, recent work \cite{hardt2016equality} shows that any definition of fairness that depends merely on the joint probability distribution is not necessarily capable of detecting discrimination. Therefore, many approaches \cite{khademi2019fairness,barabas2018interventions,kusner2017counterfactual,zhang2018equality,zhang2018fairness} are proposed to address the problem of unfairness through the lens of causality. 

In general machine learning, causal-based fairness notations are mostly defined on intervention or counterfactual. To measure the unfairness in causal-based fairness, one challenge is understanding the causal relationships that account for different outcomes. Causal graph, as a powerful tool for causal reasoning, is usually used to represent the causal relationships among variables. Given the causal graph capturing the causal relationships, many causal effects are used to measure the unfairness. For example, ATE (as Eq.\eqref{eq:ate}, also known as Total Effect \cite{pearl2009causality}) is used to measure the effect of changing sensitive attributes to the outcomes, \citet{kilbertus2017avoiding} measure the indirect causal effects \cite{pearl2022direct} from sensitive attributes to outcomes and eliminate the directed path from sensitive attributes to outcomes except via a resolving variable, where resolving variables refer to any variables in the causal graph that are influenced by sensitive attributes in a non-discriminatory way. More details of causal-based fairness notations can be found in \cite{makhlouf2020survey}. Counterfactual fairness is a commonly used definition of fairness in causal-based fairness. Counterfactual fairness is an individual-level causal-based fairness notion, which requires that the predicted outcome should be the same in the counterfactual world as in the real world for any individual \cite{kusner2017counterfactual}. The basic idea is minimizing the ATT (as Eq.\eqref{eq:att}, some references also name it as ETT \cite{glymour2016causal,pearl2009causality,li2021personalized}) conditioned on all features to receive the same probability distribution in the factual and counterfactual world. For counterfactual fairness in recommendation, the definition is given as follows \cite{li2021towards}:
\begin{mydef}
(Counterfactual Fairness in Recommendation) The counterfactual fairness is satisfied for a recommendation model if for any user $u$ with sensitive attributes $Z=z$ and remaining features $X=x$:
\begin{equation}
    P(L_z|X=x,Z=z) = P(L_{z'}|X=x,Z=z)
\end{equation}
for all $L$ and any value $z'$ attainable by $Z$, where $L$ denotes the top-$k$ recommendation list for user $u$.
\end{mydef}

In the next section, we will introduce some causal methods for fairness in recommendation.

\subsection{Causal Methods}
\subsubsection{Reweighting}\label{sec:fairness_ips}
As we mentioned before, bias is a widely accepted cause of unfairness, thus some existing work adopts inverse propensity scoring (IPS) methods to solve the bias in recommendation. For example, popularity bias will lead to item-side unfairness that popular items may obtain more exposure opportunities. The IPS approaches the biases are caused by non-randomly assigned treatment, thus use the inverse propensity to reweight the samples to remove the bias. For example, \citet{schnabel2016recommendations} consider recommendation as treatment and apply an IPS estimator in an Empirical Risk Minimization framework for learning to solve bias in recommendation. \citet{saito2020unbiased} design an IPS-based estimator for unbiased pairwise learning. \citet{wang2021combating} use a small part of unbiased data to train a propensity model and use biased data to train an IPS-based rating model. The IPS-based approaches are easy to implement but it requires an accurate propensity estimator and suffers from high variance \cite{saito2020asymmetric,wang2019doubly}. 

Although biases in data are commonly recognized as the main causes of unfairness in recommendation, the relationship between bias and fairness has not been clearly understood or discussed. More specifically, the debiasing methods are usually proposed to improve the recommendation performance by removing bias, thus the models are evaluated by recommendation metrics instead of fairness metrics. Many works on fairness are not implemented by debiasing methods but directly designed by fairness requirements, which may result in a trade-off between accuracy and fairness. In the following part, we focus on fairness methods. More discussion of debiasing methods can be found in Section \ref{sec:unbiasedness}.

\subsubsection{Counterfactual}
Counterfactual fairness, as a causality-based definition of fairness, requires the predicted outcomes to be the same in the counterfactual world as in the factual world. To achieve counterfactual fairness,  it is important but challenging for a fair model to predict the outcomes in the counterfactual world (i.e., the sensitive attributes have been changed). \citet{ma2022learning} propose a counterfactual data augmentation module, which is trained based on a variational auto-encoder with a fairness constraint, to generate counterfactual data with different sensitive attributes. By maximizing the similarity between the representation learned from the original data and the different counterfactual data, the designed model is able to achieve counterfactual fairness. \citet{mehrotra2018towards} use counterfactual estimation to evaluate recommendation policies in terms of the trade-odd beween relevance and fairness, and propose a recommendation model considering user's tolerance towards fairness. The idea of counterfactual is used not only in fairness model design but also in fairness diagnosis. Specifically, fairness diagnosis aims to find out the reasons that cause model unfairness. Inspired by the idea of counterfactual explanation \cite{tan2021counterfactual,goyal2019counterfactual}, \citet{ge2022explainable} propose a counterfactual reasoning approach to learn critical features that significantly influence the fairness-utility trade-off and use them as fairness explanation for feature-based recommendation. 

\subsubsection{Structural Equations}
A Structural Causal Model consists of a causal graph, which captures the direct causal relations among variables, and a set of structural equations, which builds the quantitative relations among variables. As we introduced in Section \ref{sec:causalesti}, if the structural equations are given, the interventional or counterfactual outcomes can be obtained by replacing the value of variables in structural equations. Inspired by this idea, some works on fairness utilize the learned or pre-defined structural equations. For example, some works \cite{wu2018discrimination,zhang2017causal,zhang2018causal,zhang2016discrimination,zhang2016situation,zhang2017anti} model different causal effects from learned structural equations to discover discrimination and further remove them. \citet{kilbertus2017avoiding} develop a practical procedure to remove discrimination given the structural equation model. 

\subsubsection{Causal Graphs}
Some other causality-based methods utilize causal graphs to capture the underlying data generation mechanism and apply other techniques to achieve fairness. For example, \citet{huang2022achieving} use the d-separation set identified from the causal graph to design a fair upper confidence bound bandit algorithm for online recommendation. \citet{li2021towards} design a model based on a causal graph to generate feature independent user representations via adversary learning. Concretely, the model trains a predictor and an adversarial classifier simultaneously, where the predictor learns the representations for recommendation and the classifier minimizes the predictor's ability to predict the sensitive features.

\subsection{Open Problems}
As we introduced above, researchers start to realize the importance of considering causality-based fairness in recommendation \cite{hardt2016equality,makhlouf2020survey}. However, the foundation of causal fairness in recommendation has not been well established. Specifically, the fairness techniques are well explored in classification tasks, however, those techniques may not be directly migrated to the recommendation problem even if the recommendation can be considered as a classification task in some cases. For example, a straightforward method \cite{kusner2017counterfactual} to achieve counterfactual fairness in classification is removing sensitive attributes from the input to guarantee the independence between the outcomes and the sensitive features. However, in recommender systems, some existing approaches do not use features for recommendation, such as most collaborative filtering based models \cite{ekstrand2011collaborative}, but still suffer from unfairness. The reason is that the interaction information contains the hidden relationship between sensitive features and user-item interaction, and this underlying relationship will be captured by the model during the collaborative learning thus leading to unfairness. Therefore, it is critical to have more explorations about the underlying causal mechanism of unfairness. Additionally, it can help the community to establish a connection between bias and fairness.

\section{Causal Robustness in Recommendation}\label{sec:robustness}
Recently, the robustness of machine learning become increasingly important. Because model time is very time-consuming therefore, the recommender system models are not re-trained frequently in practice. Traditionally, the recommender system assumes that the pattern of the training dataset and the test dataset are the same. However, there is a difference between the training dataset and the real-world data. The difference might be caused by the naturally distribution shift or intend attacking \cite{ovaisi2022rgrecsys}.Training on such a training dataset will result in performance decreasing when we apply the model to real-world data.  In this case, how to construct a robust model is very important.

\subsection{Problem Introduction}
To begin with, we need to know which aspects harm the robustness of the recommender system.  In general the dataset will be split into three subset in the training progress (training set, validation set and test set). Most of the robustness happen on training set and test set. For example, if the training dataset if not big enough can this may cause the overfitting or underfitting problem. In this case, we may get a bad results on the test set. Specifically, the robustness problem can be categorized as following:
\begin{itemize}
\item \textbf{Distributional shift} Many exist recommender systems assume that the distribution for the training set and test set are identical. However this assumption do not meet the real-world scenarios, and this makes lots of exist recommendation models cannot achieve the performance we expect when we deploy them online \cite{shen2021towards}. Many of the current recommender system models are trained based on existing collected datasets. The distribution of the data can be different when the new model is deployed \cite{cao2016non}. Even though the data is new collected, there are still transformation risks \cite{ovaisi2022rgrecsys}. Because there is a time period between training the model and deploying the model. It is long enough for the information of the collected be to attacked or changed due to the processing error.
\item \textbf{Transformation} Most of the recommender system are training with the users' and items' feature. Based on the feature, the recommender system can provide the users a set of recommendation items. However, the users' and items' feature can be corrupted or misleading. For example, if users use VPN when they purchase a item, then the location information could be wrong. With such data, the recommender system may provide wrong items to the users.

\item \textbf{Attack} Except the transformation, attack may also change the correctness of the training data. With the development of e-commence, more people start to purchase items online. There are huge benefits associated with this. Therefore, the system are highly likely to be attacked with the purpose to increase or decrease the ranking of some items. For example, the users may unwillingly changed their rating and reviews of the purchased product.

\item \textbf{Sparsity} Compared to the huge number of user and item, most of the sequential recommendation train set is sparsity. As we mentioned above, this may cause overfitting or underfitting problem and lead to low performance on test set.
\end{itemize}

\subsection{Method in Robustness}
\subsubsection{Counterfactual}
Recommendation model suffers from the data sparsity problem. For example, in the e-commerce application, compared to the large number of user and items, the users purchase history is quiet sparsity. Therefore, the model cannot get enough data for training and result in the low prediction performance. One approach to solve this problem is counterfactual data augmentation. By using counterfactual reasoning to generate new training data, and together with the original data can enhance the performance of the recommendation model. Counterfactual Data-Augmentation Sequential Recommendation (CASR) provides us a framework to solve the problem \cite{wang2021counterfactual}. For a training sample ($\{u,t^1,t^2,...t^l\},t^{l+1}$), the model will first indicate an index d, and replace a $t^d$ with an item $t^a$. Suppose $\textbf{e}_t \in R^D$ is the embedding of item t, D is the embedding size of the item. For a given sample ($\{u,t^1,t^2,...t^l\},t^{l+1}$), the model will optimize the following object:
\begin{equation}
\begin{split}
    \min_{t^a \in C}\| \textbf{e}_{t^a} - \textbf{e}_{t^d} \|^2_2 \;\;\;\;\;\;\;\;\;\;\;\;\;\;\;\;\;\;\;\;\\
    s.t. \; t^{l+1} \neq arg \max_{t \in I} \mathcal{S}(t|u,t_1,...,t^{d-1},t^a,t^{d+1},...,t^l)\\
\end{split}
\end{equation}
where $\mathcal{I}$ is the set of all item.$\mathcal{C}$ is the item set for replacement, which can be specified as $\mathcal{I}$ or other set to involve of some prior knowledge. $\mathcal{S}$ is the sampler used to generate new sequential data. In this function, the object tries to minimize the distance between the original item and the replace item. And the constraint make sure the changed item is not the original one. In this case, we can generate data that not the same as the original one but similar to the original one.

\subsubsection{Causal Graph}
Some existing works introduced the idea of causal representation learning to mitigate the distributional shift problem. The model split the user features into the observed group and unobserved group and set two types of preference depending on whether it is affected by the observed feature or not \cite{wang2022causal}. According to the casual graph, a framework is created to model the interaction generation procedure. And to deal with the unobserved feature, they design a new Variational Auto-Encoder (VAE) to infer the unobserved feature from the historical interaction and observed features.

\subsubsection{Reweighting}

Moreover, reweighting methods have been introduced to improve the robustness of recommendation, for example, \citeauthor{li2022causal}\cite{li2022causal} consider to enhance the robustness of recommendation when there are agnostic distributional shifts between training data and testing data. To this end, the paper introduces a personalized feature selection method for Factorization Machines (FMs) through referring to the confounder balancing approach to balance the confounders of each feature. In specific, considering there is usually no prior knowledge of the causal structure of input variables in FMs, the paper considers to treat every feature as a treatment variable and aims to estimate its causal effect on the outcome. When one feature is treated as a treatment, the other features are considered as confounders. The paper refers to the idea of confounder balancing \cite{shen2018causally, kuang2018stable} to learn a weight matrix to reweight each sample through balancing the distributions of confounders across different treatment features, so that FMs will assign a weight to each feature that implies its causal effect on the target variable, and thus help to select causal features for achieving robust recommendations.

\subsection{Open Problems}
Existing works on robustness problem can only focus on some specific problems. For example, using counterfactual data augmentation problem to mitigate sparsity problem and using causal representation learning to solve distribution shift problem. And if we apply these methods on other robustness problems, the experimental performance might decrease a lot. And most of the current existing works are unexplained. They might have good performance on solving some problems, but they cannot explain which part of the model improves the performance and which part of the model can have more improvement. An explained robustness method that can be applied on multiple problems is a great challenge.

\section{Uplift-based Recommendation}\label{sec:uplift}
Modern recommender systems usually aim to recommend items that users are most likely to interact with (e.g., click, purchase, etc.). However, users may interact with some items even without recommendation. Based on this fact, some existing works propose to recommend items with high interaction probability lift instead of high interaction probability values.

A closely related area is uplift modeling, which refers to the techniques used to estimate the incremental impact of a treatment on the outcomes. Uplift modeling is both a causal inference and a machine learning problem \cite{gutierrez2017causal}. It is a causal inference problem because the two required outcomes (i.e., receive treatment or no treatment) for calculating the incremental impact are exclusive for an individual. It is also a machine learning problem because it needs models to predict reliable uplift values for decision making. Theoretically, uplift modeling aims to estimate a treatment effect on outcomes \cite{gutierrez2017causal}. There are three main approaches in existing literature: the Two-Model approach trains two models on treated data and controlled data respectively, and uses the difference between two predictions to calculate the uplift value \cite{radcliffe2007using}; the class transformation approach builds the connection between the treated group and the controlled group based on some assumptions \cite{jaskowski2012uplift}; the direct estimation approach designs a model to directly estimate the uplift value \cite{radcliffe2011real,rzepakowski2012decision}. 
\subsection{Problem Introduction}
Recommender systems have been employed in several industrial domain to increase the profit of the business and improve user engagement. To achieve this goal, most of the recommendation models are designed to increase user action (e.g., click, purchase, etc.) by recommending items that have highest interaction probabilities. However, most recommender systems neglect a fact that users may take actions on some items regardless of whether the system recommends them \cite{sato2019uplift,sato2016modeling}. For example, a user will purchase bottled water with 95$\%$ probability and energy drink with 50$\%$ probability if the system recommends them. In the opinion of most traditional recommender systems, it would be better to recommend bottled water since it is more likely to be purchased by the user. However, if the system does not recommend them, bottled water, as a product for daily use, may still have 90$\%$ probability to be purchased, while energy drink may only have 20$\%$ probability of being purchased. Recommending energy drink seems to be a better choice since it has higher lift of purchase probability (i.e., 30$\%$ vs 5$\%$), which in turn may be expected to lead to more profit. Based on this motivation, there is a trend to design uplift-based recommender systems which aim to recommend items with high lift.

Some previous works have been aware of the impact of recommendation but have not solve it from a causal view. For example, \citet{bodapati2008recommendation} proposes a two-stage model which separately trains the awareness and satisfaction stages for items. By training the model based on firm-initiated purchase data (i.e., purchases as a consequence of recommendation) and self-initiated purchase data (i.e., purchases other than firm-initiated purchases), the model aims to recommend items that maximize the expected incremental number of purchase from recommendation. \citet{sato2016modeling} propose a purchase prediction model which incorporates individual differences in recommendation responsiveness. More specifically, the model includes user-specific and item-specific responsiveness to maximize the impact of recommendation.

An uplift in recommender systems is defined as an increase of user actions (e.g., click, purchase, like, etc.) caused by recommendations. Considering the uplift is defined as difference between situations with and without recommendation, from the perspective of causal inference, the uplift can be mathematically represented by potential outcomes. More specifically, taking recommendation as the treatment, let $Y(1)$ be the potential outcome with a recommendation, $Y(0)$ be the potential outcome without a recommendation. Considering the binary situation, $Y(1)=1$ and $Y(0)=1$ imply that a user will take actions on the item with and without recommendation, respectively. The uplift of an item for a user is $Y(1)-Y(0)$. In the following subsection, we will introduce some existing works on uplift-based recommendation models with causal inference.

\subsection{Causal Methods}
\subsubsection{Data Processing}
One challenge of estimating the uplift value is that each individual cannot observe both the factual and counterfactual outcomes (i.e., outcomes with and without recommendations). Thus there is no observed ground truth for the uplift value (i.e., the causal effect of recommendation). To overcome this issue, one possible solution is regarding the training data. \citet{sato2019uplift} propose a sampling method on the observational data for an uplift-based optimization. Specifically, by observing purchase and recommendation logs, for a given user, an item can be either purchased or not purchased and either recommended or not recommended. The proposed optimization samples positive and negative instances that are specific to the uplift task from four classes items (i.e., recommended and purchased, recommended and not purchased, not recommended and purchased, not recommended and not purchased). Therefore, by taking the sampled labels for uplift task as the ground truth, the proposed optimization is able to learn the uplift value for user-item pairs. Except for the sampling methods on observational data, training on experimental data is also an available option. 
\citet{shang2021partially} propose a reinforcement learning based approach, which incorporate a deep uplift network to learn the causal effect of different actions as a reward function. The uplift network learns from the training data collected from a randomized experiment.

\subsubsection{Counterfactual}
According to the calculation of the uplift value, a straightforward way is to estimate the counterfactual outcomes. Although the randomized experiment is ideal for estimating the causal effect, it is impractical to apply randomized experiment in all recommendation scenarios since it is time-consuming and expensive. Therefore, it is essential to estimate the counterfactual outcomes only based on the observational data. Inspired by the idea of collaborative filtering, \citet{xie2021causcf} believe that similar users have both similar tastes on items and similar treatment effect under recommendations. The proposed approach is designed based on tensor factorization with three dimensions as user, item and treatment. More specifically, for a three dimentional tensor with $m$ users, $n$ items and $l$ treatments, the element $y_{u,i,t}$ can be predicted as follows.
\begin{equation}
    \hat{y}_{u,i,t} = p_u^Tq_i + p_u^Td_t + q_i^Td_t
\end{equation}
where $p_u$, $q_i$, $d_t$ are latent representation of user $u$, item $i$ and treatment $t$, respectively. The predicted value of $\hat{y}_{u,i,t}$ is used to infer the potential outcome for a user-item pair $(u,i)$ under treatment $t$. Taking binary treatment setting as an example, the uplift value for a user-item pair $(u,i)$ can be estimated by $\hat{y}_{u,i,t=1}-\hat{y}_{u,i,t=0}$. \citet{sato2021causality} apply a matching estimator \cite{stuart2010matching} to estimate unobserved counterfactual outcomes and further estimate the causal effect for recommendation. More specifically, following the neighborhood methods in recommender systems, the proposed approach replaces the potential outcomes with the weighted average over the observed outcome for a set of neighbors to calculate the causal effect, where the neighbors can be neighborhood users or neighborhood items.

\subsubsection{Reweighting}
Estimating causal effect from the observational data only is challenging, since the ground truth is unobservable and the estimation is prone to the biases in the observational data. To overcome this issue, some existing works design IPS-based approaches to estimate unbiased causal effect for recommendation or evaluation. The unbiased estimation of the uplift value (i.e., the causal effect) can be formuted by IPS \cite{sato2020unbiased}.
In practice, IPS is prone to suffer from high variance issue. To tackle this problem, \citet{sato2020unbiased} apply capped inverse propensity scoring (CIPS) to train an unbiased uplift-based model; \citet{sato2019uplift} propose a unbiased estimator for uplift-based evaluation using self-normalized inverse propensity scoring (SNIPS) \cite{swaminathan2015self}; \citet{xiao2022towards} apply doubly robust technique \cite{bang2005doubly,funk2011doubly} to train an unbiased and robust model for uplift-based recommendation.

\subsection{Open Problems}

Existing works on uplift-based recommendation mainly focus on representing uplift value and estimating the causal effect using potential outcome framework. Structural causal model, as a power tool for causal inference, has rarely been used for uplift-based recommendation. 
Existing works using structural causal models are trying to estimate user's preference if the system recommends a certain item, which can be estimated by $do$-operations on designed causal graph. However, it is still not clear how to estimate the preference without recommendation using $do$-operation. First, the structural causal model requires the designed causal graph. Existing works on causal recommendation using structural causal models rarely explicitly involve the impact of recommendation into the causal graph, however for uplift-based recommendation, whether requiring a specific causal graph that explicitly depict the impact of recommendation still needs to be discussed. Secondly, mathematical representation of the preference without recommendation using $do$-operation is also a challenge. Finally, for uplift-based recommendation, if the designed causal graph and mathematical representation of preference without recommendation are decided, applying causal techniques on preference without recommendation may differ from existing works.

\section{Causal Unbiasedness in Recommendation}
\label{sec:unbiasedness}

Nowadays, recommendation algorithms have been widely used in several applications to alleviate information overloading in our daily life. Although recommender systems (RS) have obtained huge impacts in a wide range of real-world applications, it still faces many bias issues which are challenging, and if left unattended, will affect the long-term benefits of the recommender systems. Bias issues are common in RS since one nature of RS is the feedback loop. Following a generally accepted understanding \cite{chen2020bias,fan2022comprehensive}, the feedback loop in RS can be divided into three parts from a bird's-eye view: 1) the data collection part (user $\rightarrow$ data); 2) the model training part (data $\rightarrow$ model); 3) the model serving part (model $\rightarrow$ user). Different definitions of bias issues exist in each part and the whole feedback loop. We will introduce more details in the following parts.

\subsection{Problem Introduction}
As we mentioned, the bias issues exist in each part as well as the whole of the feedback loop. We will introduce the different definition of bias in the feedback loop of RS as follows:
\begin{itemize}[leftmargin=*]
    \item \textbf{Bias in Data} refers to the distribution difference between the collected data for training and the ideal test data. Typically, the training data for RS is observational instead of experimental. The user decision may be affected by several factors such as exposure mechanism of RS, thus the training distribution is different from the test distribution. Additionally, the training data may not truly represent user preference, misleading recommendation model to inaccurate prediction. We will introduce four kinds of bias in data as follows:
    \begin{itemize}
        \item \textit{Selection Bias}: Selection bias stems from users' explicit feedback (i.e., ratings). Selection bias means the observed ratings are not representative of all ratings due to users' selection. It is also referred as missing-not-at-random (MNAR).
        \item \textit{Exposure Bias}: Exposure bias usually happens in recommendations with implicit feedback. Since the information about which item the user dislikes is unavailable in observed data, the learning process will use unobserved interactions to represent negative preference. Exposure bias means unobserved interactions do not necessarily represent the user's negative preference since the users are merely exposed to a small portion of items.
        \item \textit{Conformity Bias}: Conformity Bias means that users tend to behave similarly to the others in the group, even if their behavior goes against their own judgment, which makes the feedback may not represent users' true preference.
        \item \textit{Position Bias}: Position bias is common in recommendation, especially the results are presented by a ranking list. Position bias means that users tend to interact with items in higher position in the recommendation list even if the items in higher position may not be highly relevant.
    \end{itemize}
    \item \textbf{Bias in Model} refers to the bias in the model design. Bias i not always harmful. In fact, the bias in model empower the model to achieve the ability to generalize the prediction to unobserved examples. 
    \begin{itemize}
        \item \textit{Inductive Bias}: Inductive bias represents the assumptions made the model designer to better learn the objective and to generalize beyond training data.
    \end{itemize}
    \item \textbf{Bias in Results} refers to the phenomenon that the recommendation algorithms tend to exhibit bias in recommendation results presented to users. Typically, the biases in recommendation results are studied from two perspectives, one is popularity bias and the other is unfairness. We have introduced fairness and related methods in section \ref{sec:fairness}, thus in this section, we will limit the bias in results to popularity bias.
    \begin{itemize}
        \item \textit{Popularity Bias}: Popularity bias refers to the phenomenon that popular items are recommended more frequently than their popularity warrant.
    \end{itemize}
    \item \textbf{Feedback Loop Bias} refers to the amplified bias introduced by the whole RS feedback loop mechanism. Data bias will lead to data imbalance and result in bias issues in recommendation results, while the biased recommendation will in turn impact the user's behavior and further amplify the bias in the future recommendation. Taking popularity bias as an example, the popular items get more exposure in the observed data, which in turn obtain increase opportunity to be recommended, resulting in amplified bias, where popular items become more popular and non-popular items become even less popular \cite{chaney2018algorithmic,jannach2015recommenders,mansoury2020feedback}. These amplified bias caused by feedback loop, if left unattended, will result in echo chambers \cite{ge2020understanding,xu2022dynamic} or filter bubble \cite{allen2017effects,chitra2020analyzing,gao2022cirs,nguyen2014exploring}, which will decrease the diversity and increase the homogenization.
\end{itemize}

In general, there are two ways for debiasing in recommender systems, one is debiasing during training and the other is debising during evaluation. Introducing causal inference into debias recommendation makes a great success in recent years. 
In the following part, we will introduce existing works on debiased recommendation models based on causal inference.

\subsection{Causal Methods}
\subsubsection{Data Processing}
To address the bias problem in recommender systems, one straightforward solution is to leverage unbiased data \cite{rosenfeld2017predicting,bonner2018causal,yuan2019improving,liu2020general,jiang2019degenerate,yu2020influence,chen2021autodebias,lin2021transfer}. As we mentioned in section, combining the limited experimental data and observational data is a possible solution under the relaxed ignorability assumption. In recommender systems, the experimental data, which is also called as unbiased data, intervene the system by using a random recommendation policy instead of a normal recommendation policy. More specifically, for each user, they do not use recommendation models to show items, but instead randomly select some items to show. Leveraging unbiased data helps to achieve debiased prediction because applying random recommendation will break the feedback loop. The key challenge is how to incorporate a small portion of unbiased data into model design. For example, \citet{rosenfeld2017predicting} and \citet{bonner2018causal} apply two recommendation models for biased data and unbiased data respectively and connect two models by regularization. \citet{yuan2019improving} learn a imputation models with unbiased data for ad click prediction. \citet{chen2021autodebias} leverage unbiased data by meta-learning. Despite the effectiveness on handle biases by using unbiased data, collecting unbiased data will randomly recommend items to users instead of using personalized recommendation model, which will inevitably hurt users’ experience and revenues of the platform.

\subsubsection{Reweighting}
Another commonly used method is based on reweighting, which use inverse propensity scores to reweight the data sample for different bias issues, such as selection bias \cite{schnabel2016recommendations,wang2021combating}, exposure bias \cite{yang2018unbiased,saito2020unbiased,zhu2020unbiased,mcinerney2020counterfactual}, position bias \cite{agarwal2019general,agarwal2019estimating,ai2018unbiased,chen2021adapting,joachims2017unbiased,qin2020attribute}, etc. The key challenge is how to estimate the propensity scores and how to apply it into optimization. Some works \cite{saito2020unbiased,lee2021dual} use popularity-based propensity estimator. Some works \cite{ai2018unbiased,joachims2017unbiased,qin2020attribute,zhu2020unbiased} propose a dual problem to both optimize a propensity estimator and a recommendation model. Some works \cite{chen2021adapting} propose to learn propensity scores from the observational data. Some works \cite{wang2019doubly,xiao2022towards} use doubly robust model to handle inaccurate propensity estimators. Inverse propensity scoring methods have some limitations, such as inaccurate propensity scores and suffering from high variance problem \cite{saito2020asymmetric}.

\subsubsection{Causal Adjustment}
Causal adjustment is another promising direction for addressing bias issues \cite{zhang2021causal,xu2021deconfounded,xu2022dynamic,wang2021deconfounded,xu2021causal}. With the help of do-operator, the designed models aim to estimate the causal preference $P(Y|U,do(V))$ with intervening item exposure rather than the pure associative preference $P(Y|U,V)$ estimated by traditional recommendation models. 
Intuitively, it can be understood as to answer a counterfactual question: what would the preference be if we intervene to expose the item to the user? Causal adjustment is used to estimate the causal preference with observational data. More specifically, causal adjustment includes back-door adjustment \cite{glymour2016causal}, front-door adjustment \cite{glymour2016causal}, etc. Based on the designed causal graph representing the underlying mechanism of data generation in recommender systems, the first thing is to identify a set of variables satisfying the corresponding criterion (e.g., back-door criterion for back-door adjustment, front-door criterion for front-door adjustment), then apply causal adjustment on identified variable set to estimate the causal preference. For example, \citet{zhang2021causal} apply back-door adjustment to mitigate the exposure bias caused by the item popularity; \citet{xu2021deconfounded} leverage front-door adjustment to remove the effect of unobserved confounders; \citet{wang2021deconfounded} utilize back-door adjustment to mitigate the effect of popularity bias. Causal adjustment requires to identify a set of variables satisfying the corresponding criterion, however, given a reasonable causal graph for recommender systems, it is not always find out a set of variables satisfying such criterion. But the designed causal graph will guide the model design from other ways.
\subsubsection{Causal Graph}
Causal graph, as an effective and powerful tool for causal modeling, is used to depict the data generation process in recommender systems. Based on the designed causal graph, researchers will take it as the guidance to design causal models for debiasing \cite{zhao2021popularity,zheng2021disentangling}. For example, \citet{zhao2021popularity} and \citet{zheng2021disentangling} disentangle the effect from bias and user's preference based on the designed causal graph and recommend items solely based on user's preference; \citet{wei2021model} and \citet{wang2021clicks} represent the counterfactual world based on the designed causal graph and perform counterfactual reasoning for recommendation.

\subsection{Open Problems}
Inverse propensity scoring (IPS) is a valuable method for debiasing. 
However, the effectiveness of IPS methods highly rely on the correctness of propensity scores. How to obtain correct propensity scores is still an important yet unsolved question. Existing work usually design simple propensity estimator based on some item characteristics, such as popularity-based propensity \cite{saito2020unbiased}, or learn the propensity scores from data \cite{ai2018unbiased,joachims2017unbiased,qin2020attribute,zhu2020unbiased,chen2021adapting}. Whether using correct propensity scores can be only estimated indirectly through the improvement for recommendation performance. Therefore, quantitative evaluation of the correctness of propensity scores is still an open problem and need further exploration.

\section{Open Problems and Future Directions}\label{sec:openproblems}

\subsection{Underlying Causal Mechanisms}
Recall the existing works we introduced above, most of them are based on the underlying causal mechanisms of recommender systems, which are represented by pre-defined causal relations. In general, there are three levels of pre-defined causal relations. The first level is identifying cause and effect only. For example, IPS methods only investigate the quantitative relationship between two variable, one is cause, the other is effect. For example, in some IPS based models for debiasing in recommendation, the cause is item exposure, and the effect is the probability of interactions. The second one is defining causal graphs, which identify the causal relationship between all variable pairs (i.e., whether causal relation exists, the direction of causal relation if exists). By pre-defining the causal graph, some existing works design models with the guidance of causal graph. For example, some works \cite{zhao2021popularity,zheng2021disentangling} disentangle multiple cause on the effect based on the defined causal graph to achieve unbiasedness. The last level is structural causal models, which define not only the causal relations but also quantitative relations (i.e., structural equations take causes as input and return the value of effect). The effectiveness of proposed models are highly related to the correctness of underlying causal mechanism. Currently, most of the existing works define the underlying causal mechanism through expert knowledge. The correctness of the pre-defined causal mechanism can only be indirectly reflected by the recommendation performance. Therefore, a direct and quantitative evaluation of the defined causal mechanisms deserves for further exploration. Another observation is that different models may have different pre-defined causal mechanisms even under the same practical scenario. As such, we believe that a universal causal mechanism should be proposed.

\subsection{Causal Discovery}
Apart from concerns about the accuracy of pre-defined causal mechanisms, another limitation of pre-defined causal mechanism is that pre-defined causal mechanisms by expert knowledge are usually quite simple, which only involve few factors into consideration. However, in real-world scenario, the decision-making process (i.e., the underlying causal mechanism of recommender systems) may involve much more factors, which beyond the comprehension of domain experts. Therefore, learning causal relations from data is an important yet unsolved problem in recommender systems. 
There exist few works \cite{wang2022sequential,xu2022causal} design causal discovery methods based on continuous optimization \cite{zheng2018dags,brouillard2020differentiable} for recommendation. The learned causal mechanism can be used for explainable recommendation or be leveraged to improve recommendation. Therefore, it is a promising direction to propose causal discovery methods for recommendation. It is also a challenge to evaluate the proposed causal discovery methods for recommendation. Since there is no ground-truth causal mechanism in real-world data, the causal discovery methods in recommendation are usually indirectly evaluated by recommendation performance. To directly evaluate causal discovery methods for recommendation, one possible solution is using simulation (we will introduce it in the next section).

\subsection{Causality Driven Simulations}
Simulation is one of the most powerful approach to build environments in which the recommender systems can be measured and analyzed. Building simulation for recommendation will benefit both industry and academia. For example, for industry, simulation provide controllable environment for practitioners to analyze the objectives of interest, such as some business purpose, to accelerate the pace of application development without the ethical risks. For researchers in academia, due to the restrictions of accessibility of real-world recommender systems, some proposed methods cannot be evaluated. This issue can be addressed by using simulations. Existing simulations leverage reinforcement learning techniques to simulate the decision making process under a designed environment. However, existing simulations without underlying causal mechanisms may lead to inaccurate and unstable decision-making. Leveraging causal mechanisms into simulation will achieve more stable system for long-term analysis and causal-related analysis as well. For example, causality driven simulations can be used to evaluate causal discovery methods in recommendation. Thus, causality driven simulation will play an essential role in recommender systems, which deserves further explorations.

\section{Conclusion}\label{sec:conclusion}
In this survey, we provide comprehensive review of causal inference methods for recommendation. We first provide the fundamental knowledge of recommender systems. We then introduce existing work in perspective of both causal inference and recommender systems.  More specifically, on the one hand, we introduce knowledge about causal inference and demonstrate its connection with recommender systems, on the other hand, we introduce different problems in recommender systems and how causal inference applied. Finally, we further list some open problems and future directions. We hope this survey can benefit researchers and practitioners in this area and inspire more research work in causal inference for recommendation.

\bibliographystyle{IEEEtranN}
\bibliography{reference}
\end{document}